\documentclass[twocolumn,floatfix,amsmath,amssymb]{revtex4}
\usepackage{graphicx}
\usepackage{dcolumn}
\usepackage{color}

\newcommand{\be}{\begin{equation}}
\newcommand{\ee}{\end{equation}}

\begin{document}
\title{Passive scalars: mixing, diffusion and intermittency \\
    in helical and non-helical rotating turbulence}

\author{P. Rodriguez Imazio$^{1,2}$ and P.D. Mininni$^1$}
\affiliation{$^1$ Departamento de F\'{\i}sica, Facultad de Ciencias
                  Exactas y Naturales, Universidad de Buenos Aires and 
                  IFIBA, CONICET, Cuidad Universitaria, Buenos Aires 
                  1428, Argentina.\\
                  $^2$ Laboratoire de Physique Statistique, Ecole
                  Normale Sup\'erieure, CNRS, 24 rue Lhomond, 
                  75005 Paris, France.}
                 
\date{\today}

\begin{abstract}
We use direct numerical simulations to compute structure functions,
scaling exponents, probability density functions and turbulent
transport coefficients of passive scalars in turbulent rotating
helical and non-helical flows. We show that helicity affects the
inertial range scaling of the velocity and of the passive scalar when 
rotation is present, with a spectral law consistent with 
$\sim k_{\perp}^{-1.4}$ for the passive scalar variance spectrum. 
This scaling law is consistent with the phenomenological argument
presented in \cite{imazio2011} for rotating non-helical flows, 
wich states that if energy follows a $E(k)\sim k^{-n}$ law, then the 
passive scalar variance follows a law $V(k) \sim k^{-n_{\theta}}$ with 
$n_{\theta}=(5-n)/2$. With the second order scaling exponent obtained
from this law, and using the Kraichnan model, we obtain anomalous
scaling exponents for the passive scalar that are in good agreement 
with the numerical results. Intermittency of the passive scalar is 
found to be stronger than in the non-helical rotating case, a result 
that is also confirmed by stronger non-Gaussian tails in the 
probability density functions of field increments. Finally, Fick's 
law is used to compute the effective diffusion coefficients in the 
directions parallel and perpendicular to the rotation
axis. Calculations indicate that horizontal diffusion decreases in 
the presence of helicity in rotating flows, while vertical diffusion 
increases. We use a mean field argument to explain this behavior 
in terms of the amplitude of velocity field fluctuations.
\end{abstract}

\maketitle

\section{Introduction}

The study of passive scalar advection, mixing and diffusion by 
anisotropic turbulence has gained more and more relevance over 
the years. Nowadays, it is well known that passive scalars 
share similarities with three-dimensional Navier-Stokes 
turbulence \cite{Sreenivasan,Warhaft}, presenting a direct cascade, 
anomalous scaling and intermittency \cite{Kraichnan,Falkovich}.
Moreover, the study of passive scalar mixing in turbulent anisotropic
flows is of interest in a wide variety of geophysical and
astrophysical problems, such as the transport of chemical elements 
in rotating stars \cite{Schatzman,Charbonnel,Rudiger}, the 
geodynamo \cite{Roberts}, vertical transport and diffusion in the 
oceans \cite{Rotunno,Osborn}, and the transport of polutants 
and aerosols in the atmosphere \cite{Csanady}. 

Turbulent transport of passive scalars in rotating flows was 
previously studied in \cite{Branden,imazio2011,imazio2013}, 
although it has received less attention than the transport of 
passive scalars in isotropic turbulence \cite{Warhaft,Falkovich,Celani}.
Moreover, the effect of helicity in the passive scalar transport in 
rotating flows has been practically ignored so far. It is known that 
helicity plays a key role in many problems such as in the dynamo 
effect \cite{Komm08,Branden11}, and the effect of flow helicity in 
the transport of passive vectors has been the subject of study in 
astrophysics for many years \cite{Moffat70}. Results in 
\cite{Moffat83,Chkhetiani06} for isotropic turbulence indicate 
that passive scalar transport is sensitive to whether the flow is 
helical or not. In laminar flows, and in particular in biological 
flows, it has been found that helicity enhaces transport and 
mixing \cite{Goldstein08}. 

As helicity affects the direct cascade of energy in rotating flows 
\cite{mininni09}, leading to a steeper energy spectrum, it is to be 
expected that the passive scalar cascade to smaller scales should 
also be affected by the presence of helicity (see, e.g.,
\cite{proceeding2013}). From this point two questions naturally
arise, which this work tries to answer: Is intermittency and the 
anomalous scaling of the passive scalar changed by the presence 
of helicity? And how is the transport and mixing of the passive 
scalar affected? While the former question can be answered by 
computing scaling exponents for rotating flows with and without 
helicity, the latter requires quantification of the turbulent
transport in directions parallel and perpendicular to the rotation 
axis.

The aim of this paper is then to characterize the turbulent 
scaling, transport and diffusion of passive scalars in rotating 
helical flows. To this end, we use data from direct numerical 
simulations of the Navier-Stokes equations in a rotating frame 
plus the advection-diffusion equation for a passive scalar. We 
use a spatial resolution of $512^{3}$ grid points in a regular 
periodic grid.

The analisys is divided in two parts. First, to study the effect of 
helicity in the turbulent scaling laws of the passive scalar, we 
calculate velocity and passive scalar spectra. We compute 
structure functions for the velocity and the scalar using an
axisymmetric decomposition, and consider the corresponding 
scaling exponents to quantify intermittency in each field. We 
also calculate probability density functions (PDFs) for velocity 
field and passive scalar increments. As for non-helical rotating 
turbulence (see \cite{proceeding2013}), we find that the passive
scalar is more anisotropic than the velocity field at small scales.
However, unlike the non-helical rotating case, the passive scalar 
variance follows a spectral law consistent with 
$\sim k_{\perp}^{-1.4}$, where $k_{\perp}$  denotes wave vectors 
perpendicular to the rotation axis. This scaling is shallower 
than the one found in the non-helical rotating case 
\cite{proceeding2013}, and is correctly predicted by a simple 
phenomenological relation for the energy and passive scalar 
variance spectral indices. The passive scalar in the presence
of helicity also becomes more intermittent than in the non-helical
rotating case.

Secondly, to study passive scalar diffusion, we compute 
effective anisotropic transport coefficients using the method used 
first in \cite{meneguzzi} for stratified flows, and later in 
\cite{imazio2013} for rotating non-helical flows. Effective 
transport coefficients are obtained by studying the diffusion of 
an initial concentration of the passive scalar, and calculated using 
Fick's law by measuring the average concentration and average 
spatial flux of the scalar as a function of time. For isotropic flows, 
we confirm that helicity increases turbulent diffusion (when compared 
with non-helical flows), in good agreement with previous studies and 
theoretical predictions \cite{Moffat83,Chkhetiani06}. In the presence 
of rotation, the overall effect of rotation (irrespectively of the
content of helicity of the flow) is to decrease horizontal diffusion,
while vertical diffusion remains approximately the same as in the 
isotropic case. Helicity further decreases horizontal diffusion, but 
slightly increases vertical diffusion (compared with the non-helical 
rotating case). The decrease in horizontal diffusion is explained 
using a simple model for turbulence diffusivity based on the 
amplitude of the small scale velocity fluctuations.  

\section{Set up and simulations}

\subsection{Equations and numerical method}

Data analyzed in the following sections stems form direct numerical 
simulations of the incompressible Navier-Stokes equations for the 
velocity ${\bf u}$ in a rotating frame, and of the advection-diffusion 
equation for the passive scalar $\theta$, given by
\begin{equation}
\partial_t {\bf u} + {\bf u}\cdot \nabla {\bf u} = -2{\bf \Omega} \times 
    {\bf u} - \nabla p + \nu \nabla^2 {\bf u} +{\bf f},
\label{eq:NS}
\end{equation}
\begin{equation}
\nabla \cdot {\bf u} =0, 
\label{eq:incomp}
\end{equation}
\begin{equation}
\partial_t {\theta} + {\bf u}\cdot \nabla {\theta} =  
    \kappa \nabla^2 {\theta} +\phi.
\label{eq:theta}
\end{equation}
Here $p$ is the pressure divided by the mass density (taken to be 
uniform in all simulations), $\nu$ is the kinematic viscosity, and 
$\kappa$ is the scalar diffusivity. Also, ${\bf f}$ is an external
force that drives the turbulence, $\phi$ is the source of the scalar 
field, and ${\bf \Omega} = \Omega \hat{z}$ is the rotation angular 
velocity. 

To solve Eqs.~(\ref{eq:NS})-(\ref{eq:theta}) we use a parallel 
pseudospectral code in a three dimensional domain of linear 
size $2\pi$ with periodic boundary conditions 
\cite{Gomez05a,Gomez05b}. The pressure is obtained by taking the 
divergence of Eq.~(\ref{eq:NS}), using the incompressibility condition 
given by Eq.~(\ref{eq:incomp}), and solving the resulting Poisson 
equation. The equations are evolved in time using a second order 
Runge-Kutta method. The code uses the $2/3$-rule for dealiasing, 
and as a result the maximum resolved wavenumber is $k_{max} = N/3$, 
where $N$ is the number of grid points in each direction. All 
simulations are well resolved, in the sense that the Kolmogorov 
dissipation wavenumbers for the kinetic energy and passive scalar 
variance, respectivelly $k_\nu$ and $k_\kappa$, are smaller than the 
maximum wavenumber $k_{max}$ at all times. More details of the 
numerical procedure can be found in \cite{imazio2011}.

\subsection{Dimensionless numbers and parameters}

We will characterize the simulations using as dimensionless 
numbers the Reynolds, Pecl\`et, and Rossby numbers, defined 
as usual respectively as
\begin{equation}
\textrm{Re}=\frac{UL}{\nu},
\end{equation}
\begin{equation}
\textrm{Pe}=\frac{\nu}{\kappa} \textrm{Re},
\end{equation}
\begin{equation}
\textrm{Ro} = \frac{U}{2L\Omega},
\end{equation}
where $U$ is the r.m.s.~velocity, and $L$ is the forcing scale of the 
flow defined as  $2\pi/k_F$ with $k_F$ the forcing wave number. In all
simulations $U$ is close to unity in the turbulent steady state, and the
kinematic viscosity is  $\nu=6 \times 10^{-4}$. The molecular scalar 
diffusivity is set equal to the kinematic viscosity for all runs,
resulting in $\textrm{Pe} = \textrm{Re}$.

\subsection{Initial conditions and external forcing}

We performed a set of non-helical simulations and a set of helical 
simulations with varying Rossby numbers (see Table \ref{table:21}).
In all cases, we first conducted a simulation solving only 
Eqs.~(\ref{eq:NS}) and (\ref{eq:incomp}) (i.e., the incompressible 
Navier-Stokes equations without a passive scalar), starting from 
the fluid at rest (${\bf u} = 0$), and applying a random isotropic 
external mechanical forcing ${\bf f}$ to reach a turbulent steady 
state. This turbulent steady state was integrated for at least $13$ 
turnover times. The mechanical forcing ${\bf f}$ used to sustain 
the turbulent velocity field was a superposition of Fourier modes 
with random phases, delta-correlated in time, with tuneable 
injection of helicity using the methods described in 
Ref.~\cite{patterson}.

The procedure described above resulted in several runs as listed in 
Table \ref{table:21}, with runs named with the letter ``A'' 
corresponding to simulations for wich the forcing inyected zero mean 
helicity, and runs labeled as ``B'' corresponding to runs with maximal 
injection of helicity. The final state of the velocity field in the
turbulent steady state of these runs was used as initial condition for 
multiple runs in which the external mechanical forcing ${\bf f}$ was 
maintained, but a passive scalar was injected either as an initial
concentration $\theta(t=0, {\bf x})$, or randomly injected in time
using the source $\phi$.

These two different ways to inject the passive scalar depended on the 
properties of the scalar that were studied. To characterize scaling
laws and intermittency of the passive scalar in rotating helical and
non-helical flows, the source term $\phi$ was used to reach a 
turbulent steady state in the variance of the scalar as well as in the
kinetic energy. To this end, the source $\phi$ was chosen as a 
superposition of Fourier modes with random phases, delta-correlated 
in time, injected at the same wavenumbers $k_{F}$ used in the 
mechanical forcing ${\bf f}$.

Instead, to study passive scalar turbulent diffusion, and to compute 
effective transport coefficients, we turned off the source term in
Eq.~(\ref{eq:theta}) (i.e., we set $\phi =0$). We then imposed two
different initial conditions for the passive scalar, and integrated
the velocity field and the passive scalar from those conditions to 
characterize horizontal and vertical diffusion. In each case, we 
used as initial condition Gaussian profiles as follows:
\begin{equation}
\theta(t=0,x_i) = \theta_0 e^{- (x_i-\mu)^{2}/\sigma^{2}},
\end{equation}
where $i=1$ or 3 (i.e., the initial profile can be a function solely 
of $x_1=x$, or solely of $x_3=z$),  $\mu =\pi$ (the profile is 
centered in the middle of the box, with the box of length $2\pi$), 
and $\sigma=1$. When $x_1=x$ is used, this allows us to study 
the diffusion of the initial profile in the direction perpendicular to 
rotation (or ``horizontal''), while when $x_3=z$ is used, we study 
diffusion in the direction parallel to rotation (or ``vertical''). For 
a few runs, we verified explicitly that the diffusion in the $x$ and 
$y$ directions was the same (to be expected as rotating flows 
tend to be axisymmetric). These runs with no forcing and with 
Gaussian initial profiles for the scalar will be labeled with a 
subindex indicating the dependence of the initial profile (e.g., 
runs labeled $A1_{x}$ or $A1_{z}$ indicate the run $A1$ was 
continued with an initial Gaussian profile for $\theta$ that 
depends respectively on $x$ or on $z$).

\begin{table}
\caption{\label{table:21}Parameters used for the simulations: 
$k_F$ is the forcing wave number, $\Omega$ is the rotation rate, 
$\textrm{Ro}$ is the Rossby number, $\nu$ is the kinematic viscosity, 
$\textrm{Re}$ is the Reynolds number, and 
$H=\left<{\bf u}\cdot \nabla \times {\bf u}\right>$ is the 
mean helicity. Note that runs labeled with ``A'' have helicity 
fluctuating around zero, while runs labeled with ``B'' have 
non-zero helicity.}
\begin{ruledtabular}
\begin{tabular}{c  c  c  c  c  c  c}  
{Run}& {$k_{F}$}&{$\Omega$} & $\textrm{Ro}$ & {$\nu$} & $\textrm{Re}$ & {$H$} \\
\hline
{A1} & {2} & {$0$} & {$\infty$} & {$6\times 10^{-4}$} & {$525$} & {$0$}\\
{A2} & {2} & {$8$} & {$0.02$} & {$6\times 10^{-4}$} & {$525$} & {$0$}\\
{A3} & {2} & {$16$} & {$0.01$} & {$6\times 10^{-4}$} & {$525$} & {$0$}\\
{B1} & {2} & {$0$} & {$\infty$} & {$6\times 10^{-4}$} & {$525$} &{$\approx2$}\\
{B2} & {2} & {$8$} & {$0.02$} & {$6\times 10^{-4}$} & {$525$} & {$\approx2$}\\
{B3} & {2} & {$16$} & {$0.01$} & {$6\times 10^{-4}$} & {$525$} & {$\approx2$}\\
\end{tabular}
\end{ruledtabular}
\end{table}

\section{Turbulent scaling laws}

In this section we present numerical results for the energy and
passive scalar spectra, structure functions, and PDFs for helical 
and non-helical rotating flows. To get the results in this section, 
the simulations in Table \ref{table:21} were continued forcing the 
velocity and the passive scalar, to reach a turbulent steady state 
in both quantities. We first present the methods used to analyze 
the data, then presents the results for the spectra and inertial 
range scaling laws, and finally we characterize intermittency 
using structure functions and PDFs. We also compare the data with
predictions from a simple phenomenological model, and from 
Kraichnan model for the passive scalar.

\subsection{\label{sect:Analisys1}Methods}

In this first part of the paper, the analisys consist on the 
characterization of flow anisotropy, scaling laws, and intermittency. 
To this end we consider power spectra, structure functions, and PDFs 
of passive scalar and velocity field increments for all runs.

As a result of the anisotropy introduced by rotation, we consider 
reduced perpendicular energy and passive scalar spectra, namely
$E(k_{\perp})$ and $V(k_{\perp})$. These reduced spectra are defined 
by summing the power of all (velocity or passive scalar) modes in 
Fourier space over cylindrical shells with radius $k_{\perp}$, 
with their axis aligned with the direction of the rotation axis.

To compute structure functions and PDFs, field increments must 
be defined first. Given the preferred direction introduced by 
rotation, it is natural to consider an axisymmetric decomposition 
for the increments. In general, the longitudinal increments of the 
velocity and the increments of the passive scalar fields are defined 
respectively as
\begin{equation}
\delta u({\bf x},{\bf l})=\left[{\bf u}({\bf x}+{\bf l})-{\bf u}({\bf
    x}) \right] \cdot \frac{{\bf l}}{|{\bf l}|}, 
\label{eq:deltav}
\end{equation}
\begin{equation}
\delta \theta({\bf x},{\bf l})= \theta ({\bf x}+{\bf l})-\theta({\bf x}), 
\label{eq:deltat}
\end{equation}
where the increment ${\bf l}$ can point in any direction. Structure 
functions of order $p$ are then defined as
\begin{equation}
S_{p}({\bf l})=\left\langle |\delta u({\bf x},{\bf l})|^p\right\rangle , 
\label{eq:S}
\end{equation}
for the velocity field, and as
\begin{equation}
T_{p}({\bf l})=\left\langle |\delta \theta({\bf x},{\bf l})|^p \right\rangle ,
\label{eq:Sp}
\end{equation}
for the passive scalar field. Here, brackets denote spatial average
over all values of ${\bf x}$.

These structure functions depend on the direction of the increment
(i.e., they do not assume any symmetry in the flow). In simulations 
without rotation, the field is isotropic and the $SO(3)$ decomposition 
is used to calculate the isotropic component of the structure
functions \cite{Arad,biferale2,biferale}. In the rotating case, due to 
the axisymmetry of the flow, we will consider only increments 
perpendicular to $\hat{z}$ (the rotation axis), and increments
parallel to $\hat{z}$. We denote the former increments using 
$l_{\perp}$, the latter with $l_{\parallel}$, and we follow the procedure
explained in detail in \cite{mininnipart2,imazio2011} to average 
over several $l_{\perp}$ directions.

This procedure to average Eqs.~(\ref{eq:S}) and (\ref{eq:Sp}) over
several directions can be summarize as follows. Velocity and 
passive scalar structure functions are computed from 
Eqs.~(\ref{eq:deltav}) and (\ref{eq:deltat}) using $26$ different 
directions for the increments $\textbf{l}$, generated by integer 
multiples of the vectors $(1, 0, 0)$, $(1,1, 0)$, $(2, 1,0)$, 
$(3, 1, 0)$, $(0, 1, 0)$, $(-1, 1, 0)$, $(-1, 2, 0)$, $(-2, 1,0)$, 
$(-1, 2, 0)$, $(-1, 3, 0)$, $(-3, 1, 0)$, $(-1, 3, 0)$ (all vectors 
are in units of grid points in the simulations), the $13$ vectors 
obtained by multiplying them by $-1$, and the two vectors 
$(0, 0, \pm 1)$ for the translations in $z$. Once all structure 
functions were calculated, the perpendicular structure functions 
$S_p(l_{\perp})$ and $T_p(l_{\perp})$ are obtained by averaging over 
the $24$ directions in the $x-y$ plane, and the parallel structure 
functions $S_p(l_{\parallel})$ and $T_p(l_{\parallel})$ can be
computed directly using the generators in the $z$ direction.

For all runs, this procedure was applied to $N_s$ snapshots of the 
velocity and of the passive scalar fields, separated by at least one 
turnover time each. For large enough Reynolds number, the structure 
functions are expected to show inertial range scaling, i.e., we expect 
that for some  range of scales $S_p\sim l_{\perp}^{\xi_p}$ and 
$T_p\sim l_{\perp}^{\zeta_p}$, where $\xi_{p}$ and $\zeta_{p}$ are, 
respectively, the scaling exponents of order $p$ of the velocity 
and scalar fields. Scaling exponents shown bellow are calculated 
for all  the snapshots analyzed in each simulation, and averaged 
over time. Errors are then defined as the mean square error; e.g., 
for the passive scalar exponents, the error is
\begin{equation}
e_{\zeta_{p}}=\frac{1}{N_s}\sqrt{\sum_{i=1}^{N_s}\left(\zeta_{p_{i}}-
    \overline{\zeta_{p}}\right)^{2}},
\label{eq:error}
\end{equation}
where $\zeta_{p_{i}}$ is the slope obtained from a least square fit
for the $i$-th snapshot, and $\overline{\zeta_{p}}$ is the mean
value averaged over all snapshots. The error in the least square 
calculation of the slope for each snapshot is much smaller than 
this mean square error and neglected in the propagation of
errors. Extended self-similarity \cite{benzi1,benzi2} is not used 
to obtain the scaling exponents.

\begin{figure}
\includegraphics[width=8.8cm]{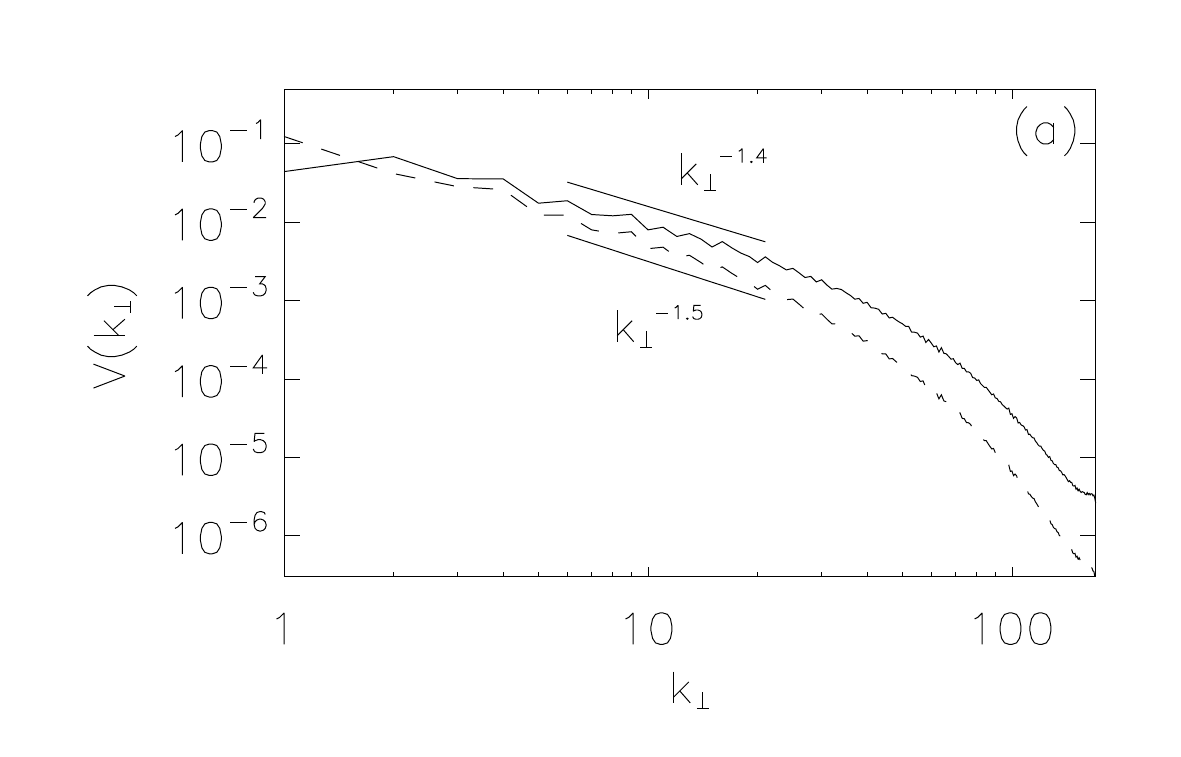}\\
\includegraphics[width=8.8cm]{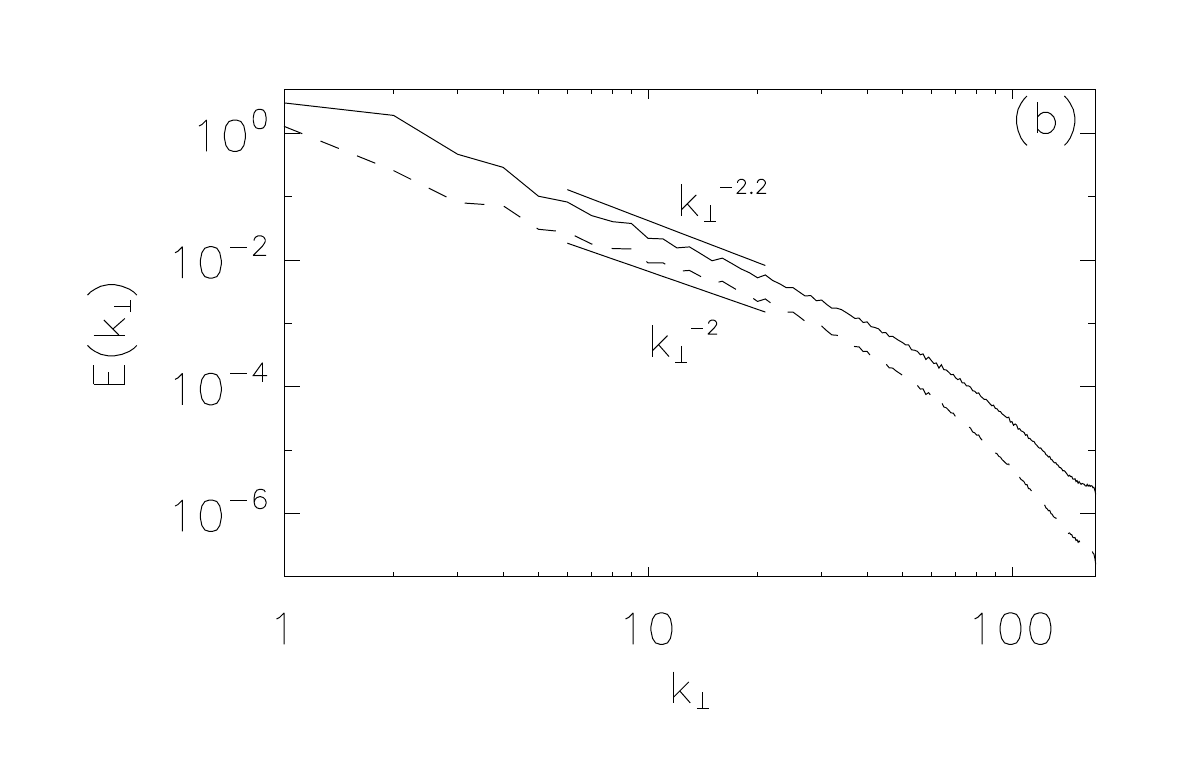}
\caption{(a) Reduced perpendicular passive scalar variance spectrum 
for run B3 (rotating and helical, solid) and for run A3 (rotating and
non-helical, dashed). (b) Reduced perpendicular energy spectrum for
run B3 (solid) and for run A3 (dashed).}
\label{fig:fig1}
\end{figure}

\subsection{Energy and passive scalar spectra}

In the presence of rotation and in the absence of helicity, the 
spectral behavior of the passive scalar is strongly anisotropic and
quasi-two dimentional \cite{imazio2011}. As previously shown in 
\cite{imazio2011}, $E(k_\perp) \sim k_\perp^{-2}$ for the velocity
field and $V(k_\perp) \sim k_\perp^{-3/2}$ for the passive scalar. 
The presence of helicity in rotating flows affects the cascade of 
energy and of the passive scalar to smaller scales. Numerical
simulations in \cite{mininni09} showed that, when helicity is present
in rotating flows, the direct cascade of helicity dominates over the
direct cascade of energy in the inertial range. This is the result of
the development of an inverse cascade of energy, which leaves less 
energy available for the system to transfer to small scales. 
Assuming the direct cascade of helicity is dominant, a spectrum 
$E(k_{\perp})H(k_{\perp}) \sim k_{\perp}^{-4}$ is obtained from
dimensional arguments \cite{mininni09}. In other words, if the
energy spectrum satisfies $E(k_{\perp}) \sim k_{\perp}^{-n}$,
then the helicity must follow a spectrum 
$H (k_{\perp}) \sim k_{\perp}^{4-n}$. As a result, the energy spectrum 
becomes steeper as the flow becomes more helical, with the limit 
of a spectral index $ n =  2.5$ for the energy in the case of a 
turbulent flow with maximum helicity (in practice, this limit cannot 
be attained since in a flow with maximum helicity the nonlinear 
term becomes negligible, resulting in no net energy transfer).

\begin{figure}
\centering
\includegraphics[width=8.5cm]{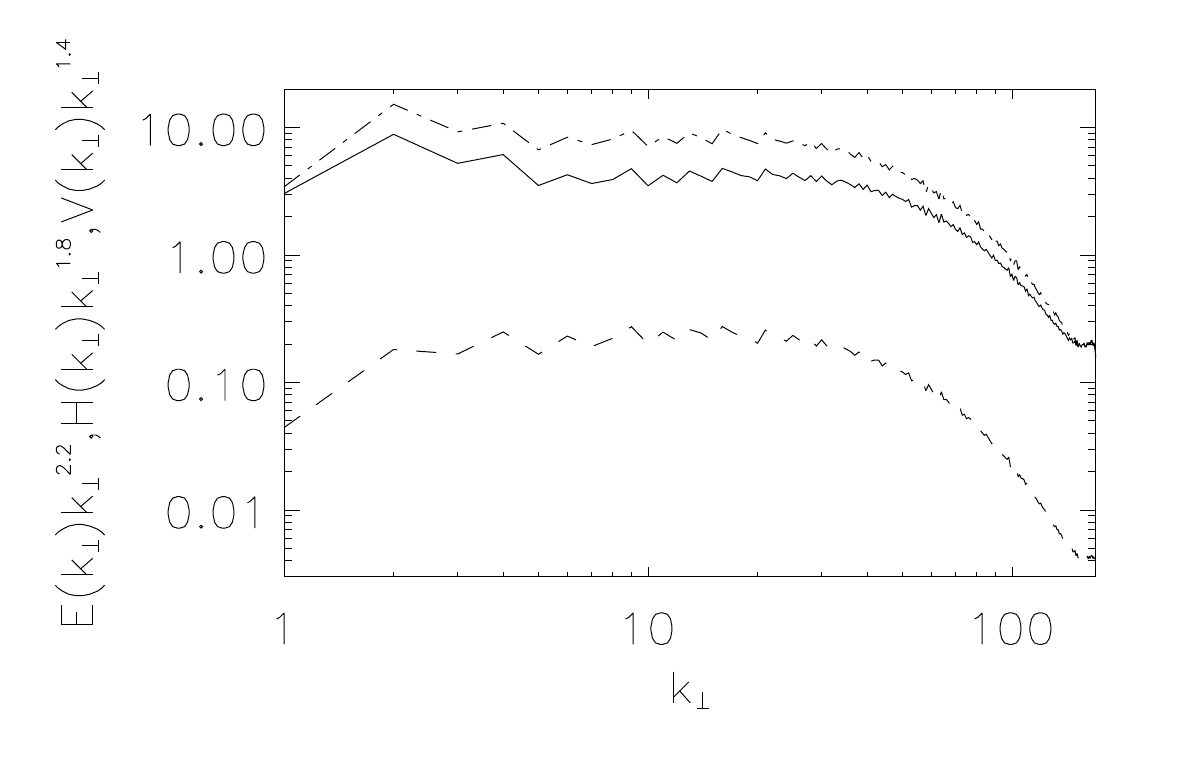}
\caption{Reduced perpendicular energy spectrum compensated 
by $k_\perp^{-2.2}$ (solid), reduced perpendicular passive scalar 
spectrum compensated by $k_\perp^{-1.4}$ (dashed), and reduced 
perpendicular helicity spectrum compensated by $k_\perp^{-1.8}$ 
(dash-dotted) in run B3 (with rotation and helicity).}
\label{fig:fig2}
\end{figure}

Figure \ref{fig:fig1} shows the energy and passive scalar reduced 
perpendicular power spectra for runs A3 and B3 (both runs with
rotation, and respectively without and with net helicity). The kinetic
energy spectrum is steeper in the presence of helicity, compatible
with $E(k_\perp) \sim k_\perp^{-2.2}$ scaling, while the passive
scalar is close to $V(k_\perp) \sim k_\perp^{-1.4}$ scaling. Although
resultion is moderate in these simulations (see \cite{imazio2011} for
more detailed studies of spectral scaling), the scaling laws can be
further confirmed in Fig.~\ref{fig:fig2}, where compensated energy and 
passive scalar spectra for run B3 are shown. In Fig. ~\ref{fig:fig2}
we also present the helicity spectrum compensated by $k_\perp^{-1.8}$
(to confirm that the helicity spectral index and energy spectral index 
add to 4). Similar scaling laws were observed in the rest of the
helical rotating runs listed in Table \ref{table:21}.

Following the phenomenological argument presented in
\cite{imazio2011}, we can explain the effect of helicity in the
scaling of the passive scalar spectrum. From Eq.~(\ref{eq:theta}), the 
passive scalar flux $\sigma$ can be estimated as 
\begin{equation}
\sigma \sim \frac {\theta_ {l_{\perp}}^{2} u_{l_\perp}} {l_\perp},
\label{eq:sigma2}
\end{equation}
If we assume that the passive scalar has a direct cascade with
constant flux $\sigma$ in the inertial range, then the passive scalar 
power spectrum $V(k_{\perp}) \sim \theta_ {l_{\perp}}^{2} / k_{\perp}$ can
be estimated, using Eq.~(\ref{eq:sigma2}), as
\begin{equation}
V(k_\perp) \sim \frac {\sigma l_\perp^{2}} {u_{l_\perp}}.
\label{eq:V1}
\end{equation}
For an energy spectrum $E(k_{\perp}) \sim k_{\perp}^{-n}$, and
therefore for a characteristic velocity at scale $l$ satisfying 
$u_{l_\perp} \sim l_{\perp}^{1-n}$, the passive scalar spectrum in 
Eq.~(\ref{eq:V1}) results
\begin{equation}
V(k_\perp) \sim \sigma l_{\perp}^{\frac{5 -n}{2}} 
    \sim \sigma k_{\perp}^{-\frac{5 -n}{2}}.
\label{eq:V12}
\end{equation}
Therefore, the spectral index for the passive scalar inertial range is 
\begin{equation}
n_{\theta} = \frac{5-n}{2}.
\label{eq:V2}
\end{equation}
This simple phenomenological argument was proposed in
\cite{imazio2011} to take into account the effect of rotation in the
spectrum of the passive scalar. Here we confirm that this argument 
remains valid in the presence of helicity in the rotating
flow. Moreover, when rotation is zero, we recover $n_{\theta}=5/3$, in 
good agreement with the Kolmogorov scaling previously observed for 
passive scalars in isotropic turbulence (see, e.g., \cite{Warhaft}).   

\begin{figure}
\centering
\includegraphics[width=8.9cm]{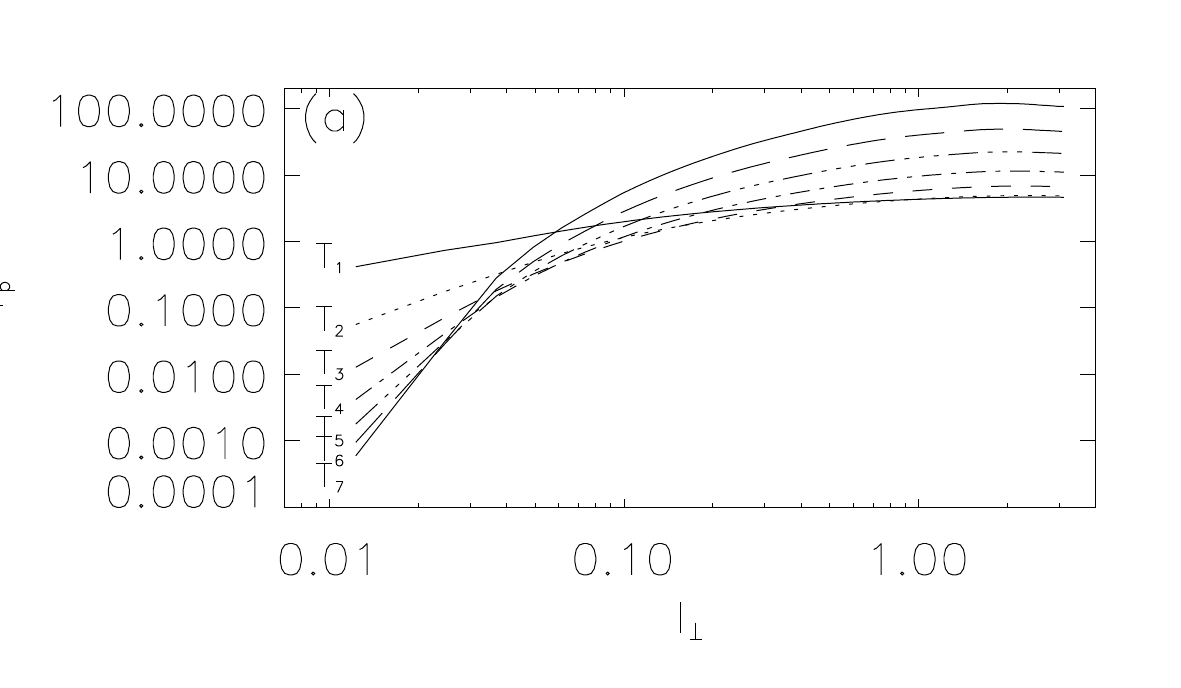}\\
\includegraphics[width=8.9cm]{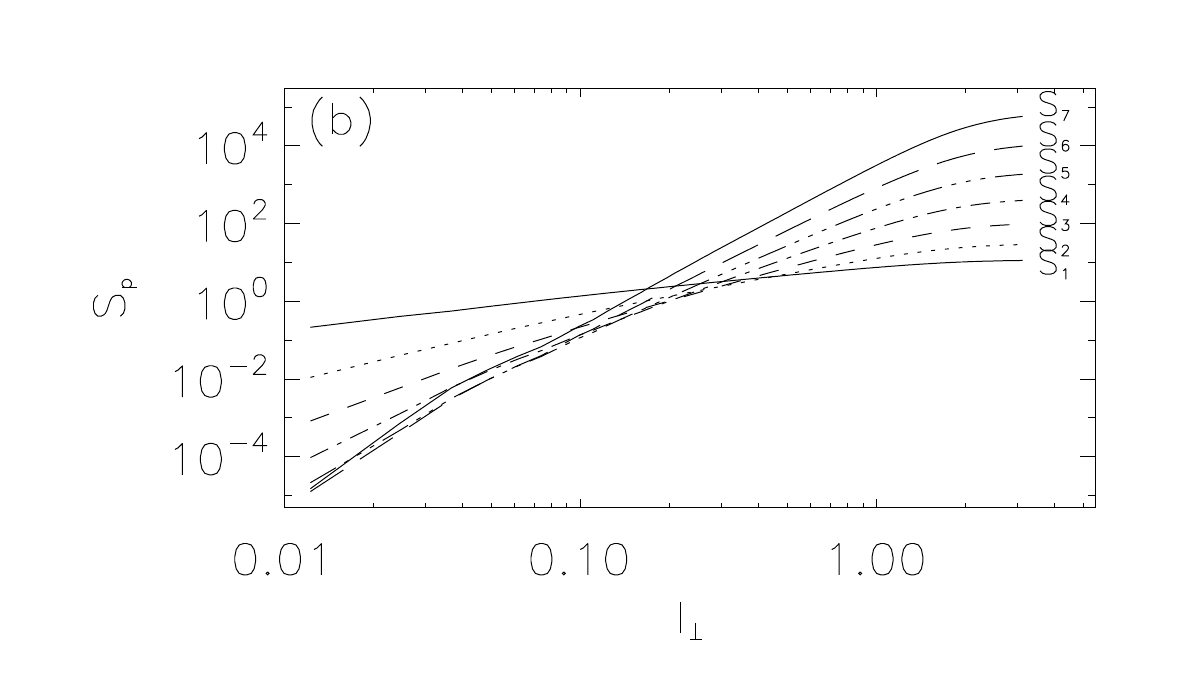}
\caption{Averaged axisymmetric structure functions (only for $l_\perp$ 
increments) up to seventh order in run B3 (rotating and helical) for
(a) the passive scalar, and (b) the velocity field.}
\label{fig:fig3}
\end{figure}

\subsection{Structure functions and scaling exponents}

Structure functions and scaling exponents for the passive scalar in 
non-helical rotating flows were studied in detail in
\cite{imazio2011}. As a result, here we focus on the simulations with
helical forcing. Figure \ref{fig:fig3} shows the axisymetric and
perpendicular (i.e., only for perpendicular increments $l_\perp$)
structure functions for the passive scalar and for the velocity field 
up to seventh order for run B3. Each curve corresponds to an average 
over $N_s= 8$ snapshots of the turbulent steady state of the 
simulation. The structure functions show a range of scales with 
approximately power-law scaling at intermediate scales, while at the 
smallest scales approach the $\sim l^{p}$ scaling expected for a 
smooth field in the dissipative range.

\begin{figure}
\centering
\includegraphics[width=8.6cm]{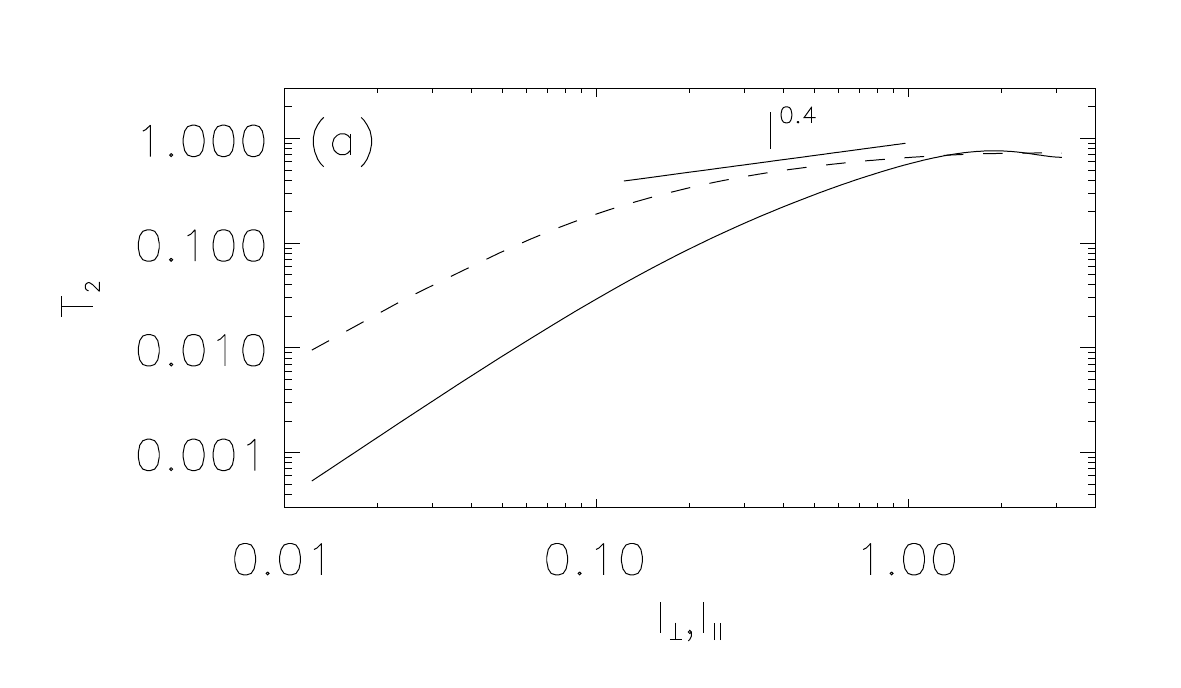}\\
\includegraphics[width=8.7cm]{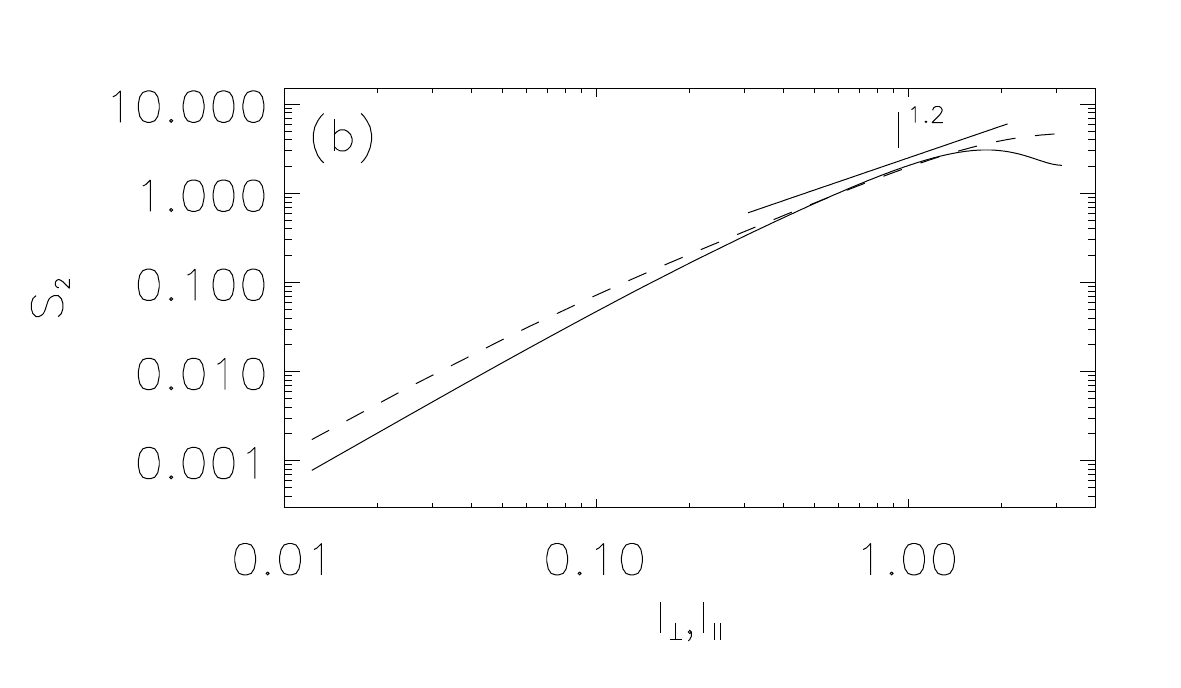}
\caption{Axisymmetric second order structure functions for run B3 
(helical with rotation) for (a) the passive scalar, and (b) the
velocity field. In both panels solid lines correspond to the
parallel structure functions, while dashed lines correspond to the 
perpendicular structure functions. Slopes indicated as references 
correspond to the time average of the scaling exponents, obtained 
from a best fit in the inertial range of the structure functions at 
different times.}
\label{fig:fig4}
\end{figure}

Figure \ref{fig:fig4} shows a detail for the same run of the passive 
scalar and velocity field second-order perpendicular structure 
functions, respectively $T_2(l_\perp)$ and $S_2(l_\perp)$, as well as 
the structure functions for increments parallel to the rotation axis, 
$T_2(l_{\parallel})$ and $S_2(l_{\parallel})$. Stronger anisotropy is 
observed at small scales for the passive scalar than for the velocity 
field, manifested as a larger difference between $T_2(l_{\parallel})$
and $T_2(l_\perp)$ than between $S_2(l_{\parallel})$ and
$S_2(l_\perp)$. Also, an inertial range with power-law scaling can 
be identified at intermediate scales in $T_2(l_\perp)$ and
$S_2(l_\perp)$. The range of scales is consistent with the wavenumbers
of the inertial range in the corresponding spectra. The slopes 
indicated as a reference in Fig.~\ref{fig:fig4} correspond to the 
time average of the second-order scaling exponents, obtained 
from a best fit in the inertial range of all structure functions at 
different times. The second-order scaling exponents (in the 
perpendicular direction) are $\zeta_2= 0.41 \pm 0.01$ for the
passive scalar, and $\xi_2=1 .22\pm0.01$ for the velocity field. 
These values are in good agreement with the spectra 
$V (k_\perp) \sim k^{1.4}$ and $E (k_\perp) \sim k^{2.2}$, wich 
from dimensional analysis lead to $T_2(l_\perp)\sim l_\perp^{0.4}$ and 
$S_2(l_\perp)\sim l_\perp^{1.2}$. From the curves in
Fig.~\ref{fig:fig3}, scaling exponents can also be computed for 
lower and higher orders. Based on the amount of statistics available, 
velocity and passive scalar exponents in the direct cascade range 
were computed for all runs up to the seventh order.

\begin{figure}
\centering
\includegraphics[width=9cm]{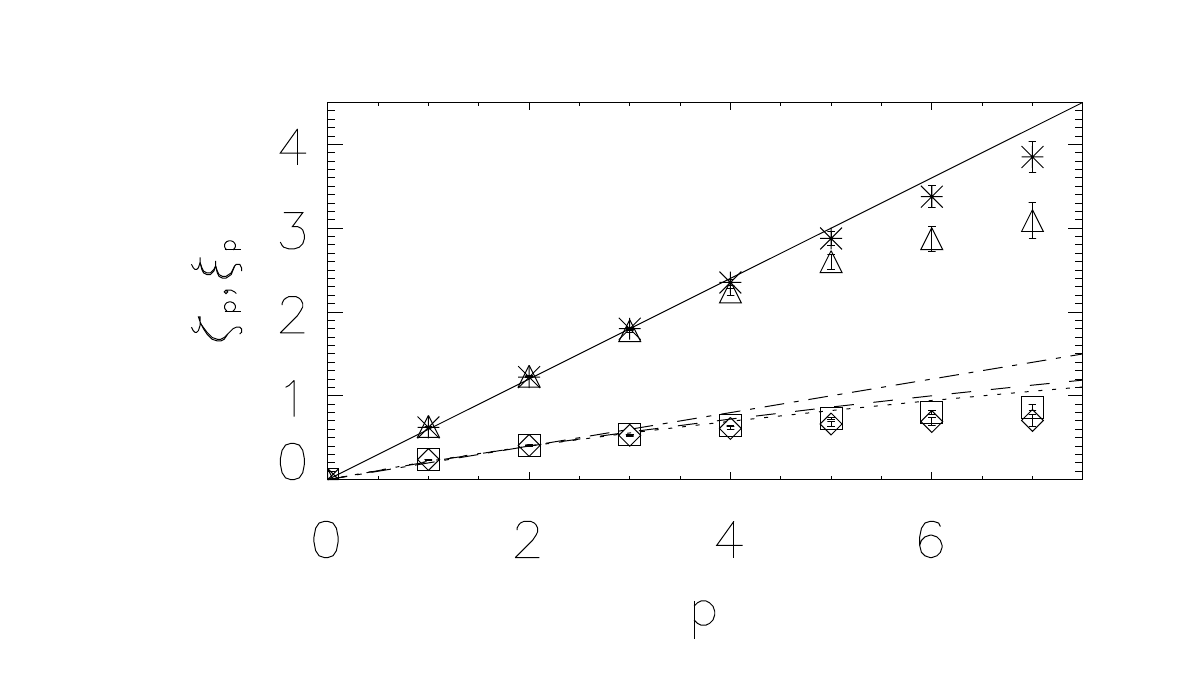}
\caption{Scaling exponents (with error bars) as a function of the
order $p$ in simulations of helical rotating turbulence, for the
velocity field (triangles for run B2 and stars for run B3, both with
helicity and with decreasing Rossby number), and for the passive
scalar (diamonds for run B2 and squares for run B3). The solid line
corresponds to the linear scaling expected for the velocity field
exponents in the absence of intermittency, while the dash-dotted line
corresponds to non-intermittent scaling for the passive scalar
exponents. The dotted and dashed lines correspond to Kraichnan's model
with $\zeta_{2}=0.4$ and respectively with $d=2$ and with $d=3$.}
\label{fig:fig5}
\end{figure}

Figure \ref{fig:fig5} shows the resulting velocity scaling exponents 
$\zeta_p$, and passive scalar exponents $\xi_p$, for runs B2 and B3 
(both helical, with $\textrm{Ro} = 0.02$ and $0.01$
respectively). Linear (non-intermittent) scalings for $\zeta_p$ and
for $\xi_p$ are shown as a reference, based on the values of the 
second-order exponents $\zeta_2$ and $\xi_2$. In Fig.~\ref{fig:fig5} 
we also show the the prediction of the Kraichnan model
\cite{Kraichnan}, which is a model for the advection and diffusion 
of a passive scalar in a random, delta-correlated in time velocity
field in a space with dimensionality $d$. The scaling exponents for
the passive scalar in this model are
\begin{equation}
\zeta_p = \frac{1}{2}\left[ \sqrt{2d\zeta_2 p+(d-\zeta_2)^{2}}+
    (d-\zeta_2)\right].
\label{eq:ansatz}
\end{equation}
For the curves in Fig.~\ref{fig:fig5}, these exponents were evaluated 
with the value of $\zeta_2$ obtained from the simulations, and using 
either $d=2$ or $d=3$.

Scaling exponents for the velocity field are similar in both runs. 
The second-order velocity field exponent is $\xi_{2}=1.22\pm0.02$ 
for run B2, and  $\xi_{2}=1.23\pm0.01$ for run B3. The velocity field 
exponents display the well-known deviations from linear scaling 
associated with intermittency, more evident for the higher order
exponents and in the simulation with larger Rossby number (i.e.,
smaller rotation rate). The deviation from strict scale invariance is
often quantified in terms of the intermittency exponent 
$\mu =2\xi_3- \xi_6$, which for these runs is $\mu = 0.6\pm0.2$ 
for run B2, and $\mu = 0.2\pm0.1$ for run B3. The decrease in the 
values of $\mu$ suggest a reduction of intermittency with increasing 
rotation, as observed before in simulations and in experiments 
\cite{Baraud03,Muller07,Seiwert08,mininni08,mininni092,mininnipart2,imazio2011}.

The passive scalar exponents for these two runs also display similar 
values.The second-order scaling exponent is $\zeta_2 = 0.41 \pm 0.01$ 
for both runs. Deviations from linear scaling are observed, and the 
intermittency exponents are $\mu_{s} = 0.36 \pm 0.06$ for run B2,
and $\mu_{s} = 0.27 \pm 0.04$ for run B3. For runs B2 and B3, 
Kraichnan's model adjusts the numerical data best with $\zeta_2 = 0.4$ 
and $d = 2$. The value of $d$ is compatible with
quasi-bidimensionalization in the spatial distribution of of the
passive scalar in the presence of rotation, as reported in the
presence of rotation in \cite{imazio2011}.

Overall, the decrease in the values of $\mu$ observed for both the 
velocity field and the passive scalar indicate a reduction of
intermittency with decreasing Rossby number. However, this reduction
is more pronounced for the velocity field than for the passive scalar.

\begin{figure}
\centering
\includegraphics[width=8.9cm]{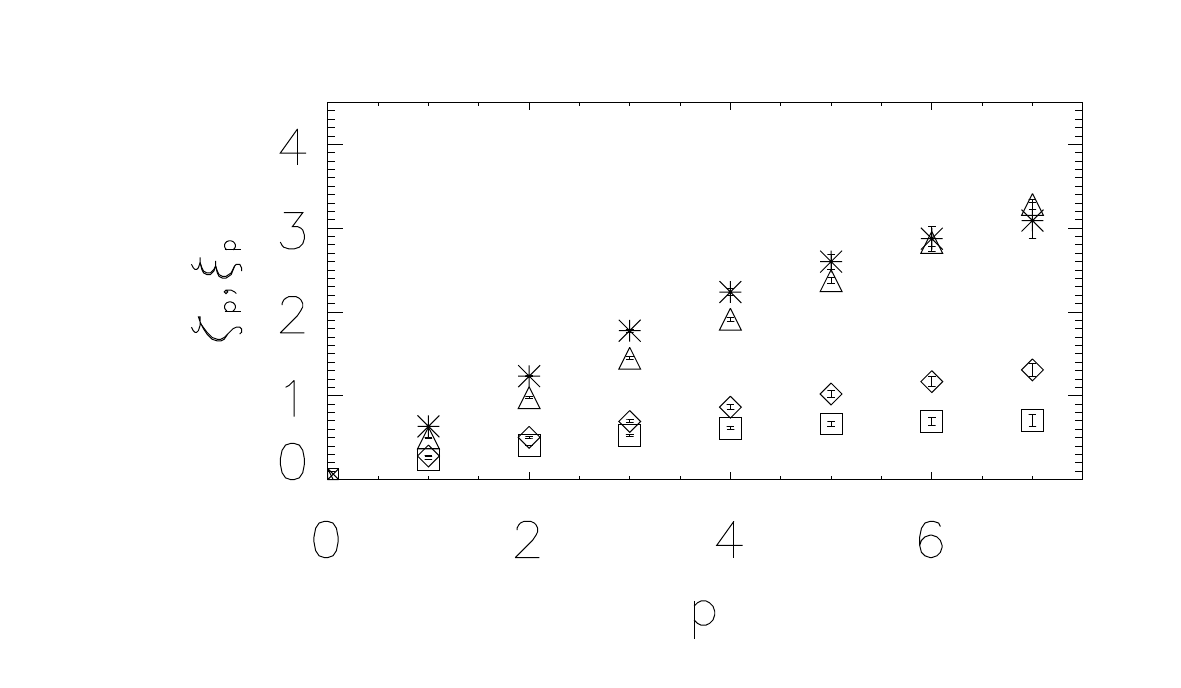}
\caption{Scaling exponents (with error bars) as a function of the 
order $p$ in simulations of rotating turbulence with and without
helicity, for the velocity (triangles for run A3 without helicity, and 
stars for run B3 with helicity), and for the passive scalar (diamonds
for run A3, and squares for run B3).}
\label{fig:fig6}
\end{figure}

Finally, we present a comparison between the scaling exponents in
rotating turbulence with and without helicity. Figure \ref{fig:fig6} 
shows the velocity field and passive scalar exponents for runs A3 and
B3 (respectively without and with helicity). Deviations from linear
(non-intermittent) scaling are larger for the passive scalar in B3,
indicating stronger intermittency in the presence of helicity.

\subsection{Probability density functions}

Intermittency and small scale anisotropy can be also studied
considering the PDFs of the field increments. In this section we
present PDFs of longitudinal increments of the $x$-component 
of the velocity field, as well as increments and spatial derivatives 
of the passive scalar concentration. Quantities shown are normalized 
by their variance, and a Gaussian curve with unit variance is shown 
as a reference.

\begin{figure}
\centering
\includegraphics[width=8.3cm]{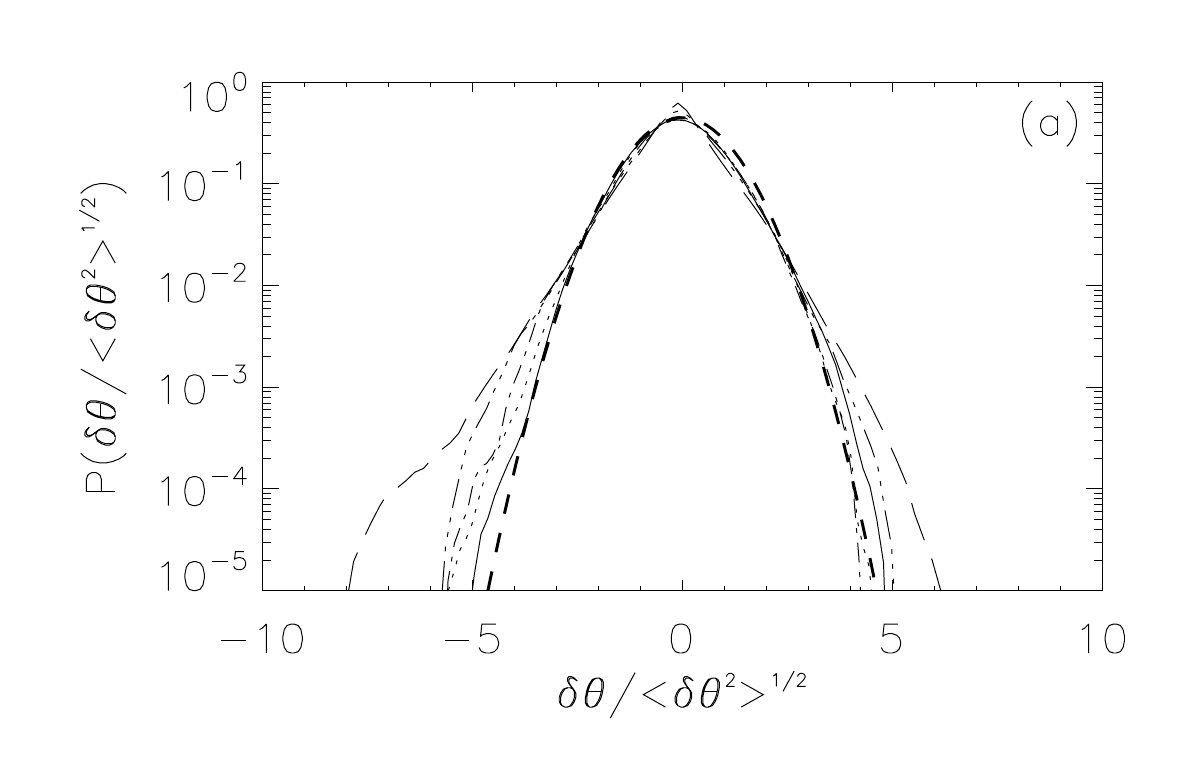}\\
\includegraphics[width=8.3cm]{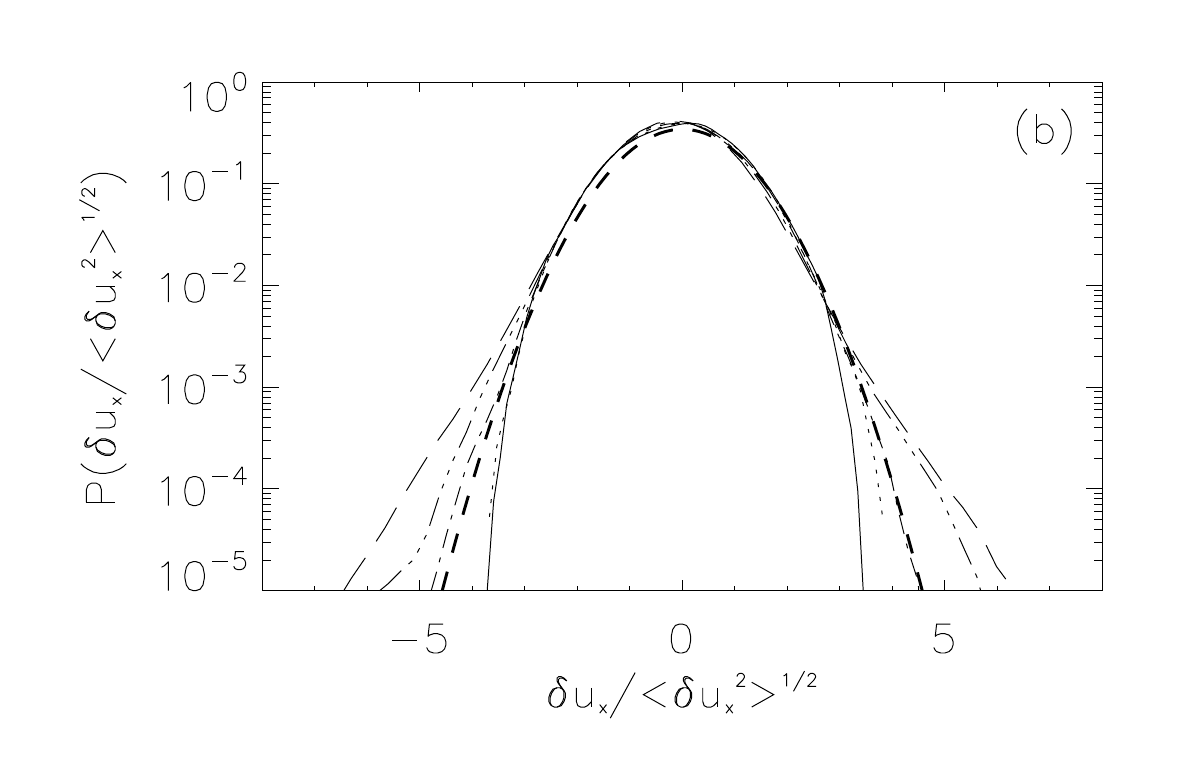}
\caption{Probability density functions in run B3, for five different 
horizontal spatial increments $l=1.6$ (solid), $0.8$ (dashed), $0.4$
(dash-dotted), $0.2$ (dash-triple-dotted), and $0.1$ (long dashes),
and for (a) the passive scalar, and (b) the $x$-component of the 
velocity field. A Gaussian curve with unit variance is indicated by 
the dotted curve. As intervals are decreased, curves depart more and 
more from the Gaussian distribution developing stronger tails.}
\label{fig:fig7}
\end{figure}

Figure \ref{fig:fig7} shows the PDFs of the velocity and of the
passive scalar increments for four different values of the spatial
increment ($l=1.6$, $0.8$, $0.4$, $0.2$, and $0.1$) in run B3 (with
helicity). All the increments were considered in the $x$-direction
(perpendicular to the axis of rotation, and for the velocity the
$x$-component was used to build longitudinal increments. As a
reference, and to compare with the values of the increments
considered, the forcing scale in this runs is $\approx \pi$, and the
dissipative scale is $\approx0.05$. Therefore, increments $l=0.8$ and
$0.4$ correspond to scales in the inertial range. The PDFs of velocity
and passive scalar increments for $l=1.6$ are close to Gaussian, while
for smaller spatial increments non-Gaussian tails develop. Note also
that in the PDFs of passive scalar increments, a strong asymmetry
develops for $l=0.4, 0.2$ and $0.1$.

\section{Turbulent diffusion}

In this second part of the paper, the aim is to characterize the 
turbulent diffusion of the passive scalar in rotating helical
turbulence, and to compare it with turbulent diffusion in non-helical
rotating turbulence, as well as with turbulent diffusion in isotropic
turbulence. To this end, we simulate the flows starting from an
initial Gaussian profile for the concentration of the passive scalar,
and we let it diffuse in directions parallel and perpendicular to the
rotation axis. We then quantify effective transport coefficients by
measuring the time evolution of the averaged concentration, and using
Fick's law.

\begin{figure}
\centering
\includegraphics[width=8.3cm]{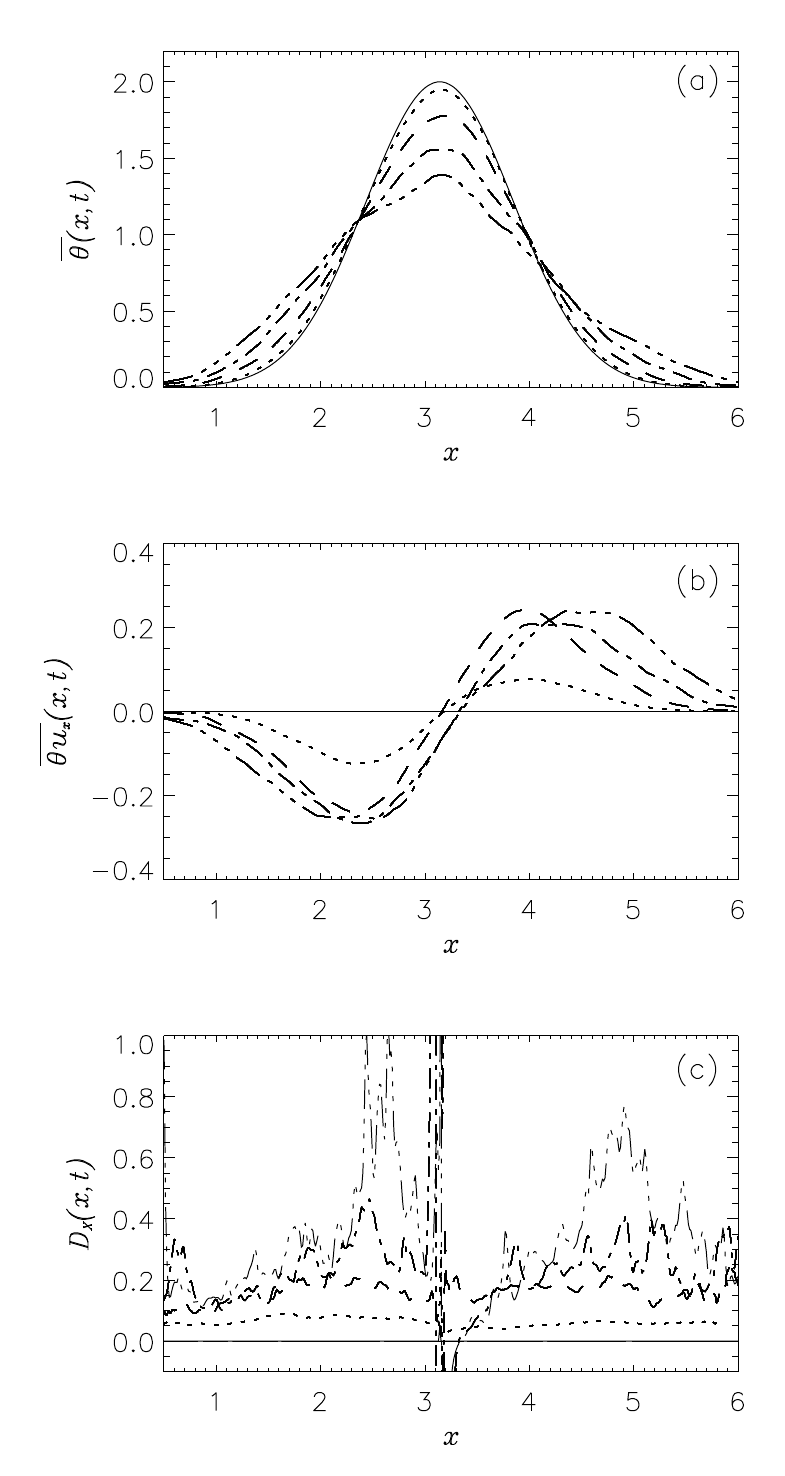}\\
\caption{(a) Averaged horizontal concentration $\overline{\theta}$ in 
run A$1_{x}$ (no rotation, no helicity) at times $t =0$ (solid),
$0.5$ (dotted), $1$ (dashed), $1.25$ (dash-dotted), and $1.5$ 
(dash-triple-dotted). (b) Horizontal flux at the same times. (c) 
${\cal D}_x(x,t)$ at the same times.}
\label{fig:fig8}
\end{figure}

\begin{figure}
\centerline{\includegraphics[width=8.3cm]{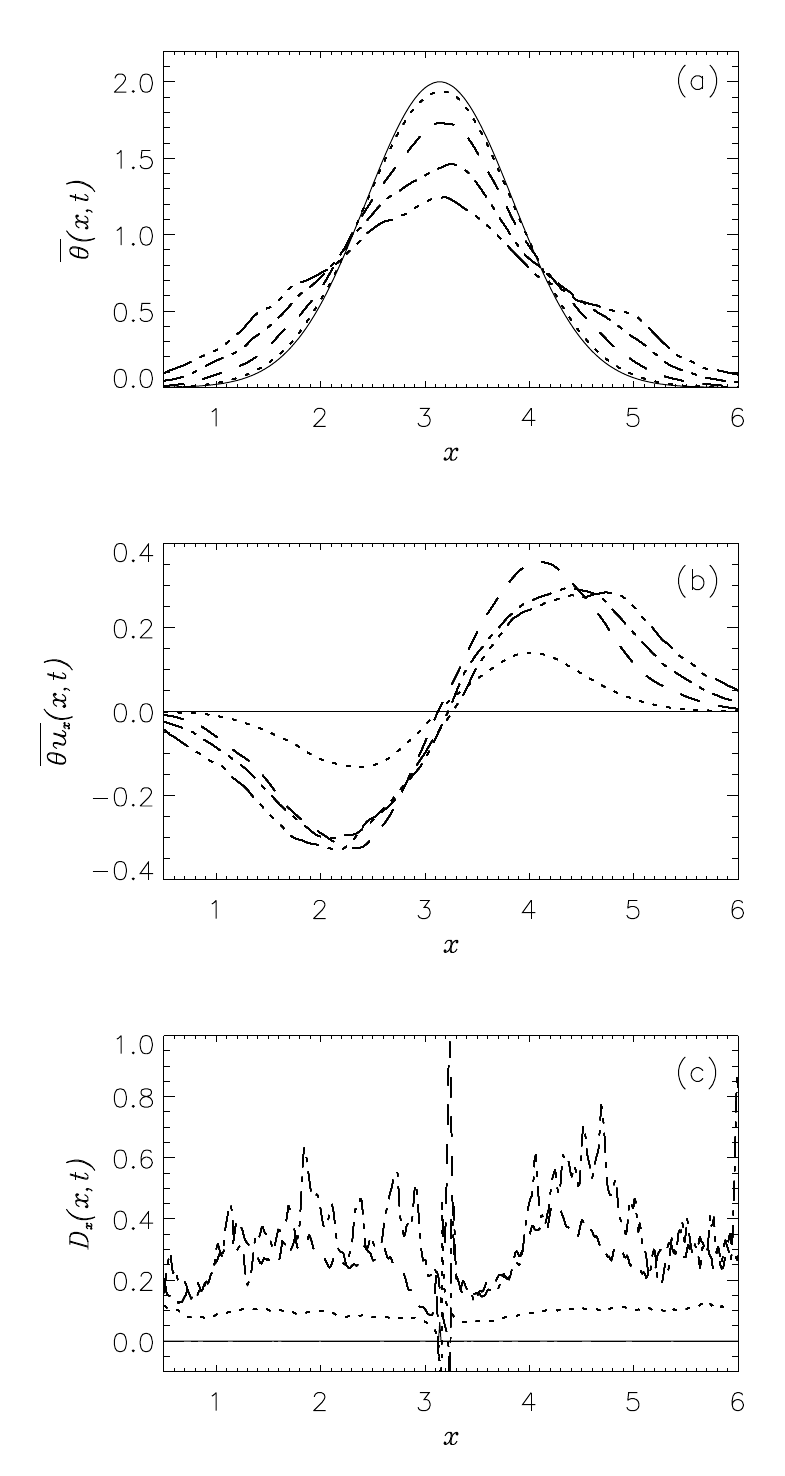}}
\caption{(a) Averaged horizontal concentration $\overline{\theta}$ in 
run B$1_{x}$ (no rotation, helical) at times $t =0$ (solid), $0.5$
(dotted), $1$ (dashed), $1.25$ (dash-dotted), and $1.5$
(dash-triple-dotted). (b) Horizontal flux at the same times. (c) 
${\cal D}_x(x,t)$ at the same times.}
\label{fig:fig9}
\end{figure}

\subsection{\label{sect:Analisys2}Methods}

Before presenting the method used to measure the turbulent diffusion,
we briefly recall how the simulations were conducted for this second
study. As in the previous section, simulations in group A (see Table
\ref{table:21}) correspond to simulations with zero mean helicity,
while simulations in group B correspond to simulations with helical
forcing and non-zero net helicity. As explained in
Sec.~\ref{sect:Analisys2}, for each run in the turbulent steady state
of the velocity field, the simulation was extended twice with the same
parameters and mechanical forcing, but with two different initial
conditions for the passive scalar: a Gaussian profile for the
concentration in the $x$-direction (to study horizontal diffusion),
and a Gassial profile in the $z$-direction (to study vertical
diffusion). To identifie these runs, an additional subindex is used in
this section to differentiate between simulations with different
dependence of the initial Gaussian profile. As examples, a run 
labeled A$1_{x}$ stands for a simulation with the parameters of run
A1 (i.e., with zero mean helicity and no rotation) and with initial 
profile of the passive scalar in the $x$-direction, while the label 
B$2_{z}$ indicates the run has helicity, rotation, and an initial
dependence of the passive scalar in the $z$-direction.

\begin{figure}
\centerline{\includegraphics[width=8.5cm]{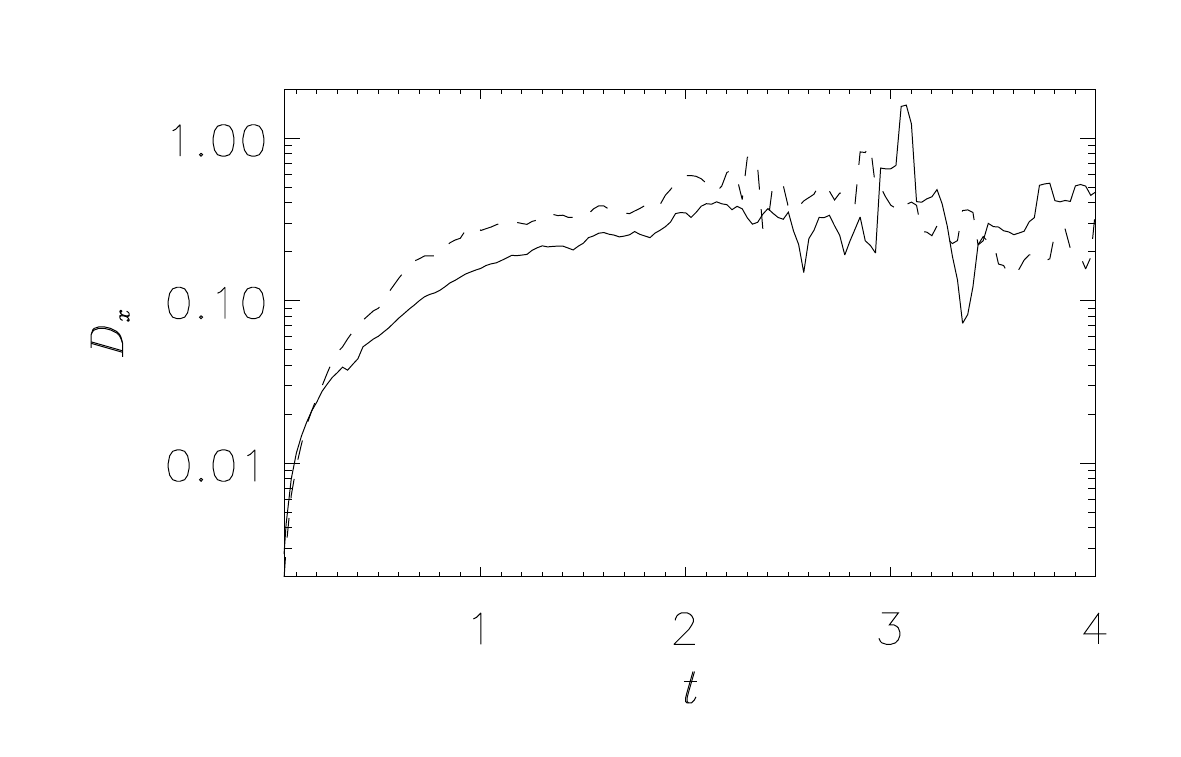}}
\caption{Horizontal turbulent diffusion as a function of time for 
runs A$1_{x}$ (solid, no rotation and no helicity) and B$1_{x}$ 
(dashed, no rotation but with helical forcing).}
\label{fig:fig10}
\end{figure}
 
In each of these runs, we let the initial profile diffuse for several
turnover times. Meanwhile, we compute and store quantities averaged 
over the two directions perpendicular to the direction over which the
original Gaussian profile varies. In particular, we consider the 
averaged passive scalar concentration $\overline{\theta}$, and the 
spatial passive scalar flux $\overline{\theta u_{i}}$, where $i=1$ or 3 
depending on the initial dependence of the Gaussian profile, and where
the averages denoted by the overbars are done over the two remaining 
Cartesian coordinates. Note the spatial flux $\overline{\theta u_{i}}$ 
represents the amount of passive scalar transported in the
$i$-direction per unit of time by the fluctuating (or turbulent)
velocity, since there is no mean flow in our simulations (we use 
delta-correlated in time random-forcing), $u_{i}$ is the fluctuating 
velocity.

Then, the pointwise effective turbulent diffusion coefficient is
given by \cite{meneguzzi}
\begin{equation}
{\cal D}_{i}(x_i,t)=\frac{\overline{\theta u_{i}}}{\partial _{x_i} 
    \overline{\theta}}.
\label{eq:DT}
\end{equation}
This coefficient corresponds to how much passive scalar is 
transported by the fluctuating velocity per unit of variation of
$\overline{\theta}$ with respect to $x_i$. As already mentioned, 
$i=1$ stands for horizontal diffusion, while $i=3$ stands for 
vertical diffusion, where the dependence on the direction of this 
coefficient is the result os the flow being anisotropic.

\begin{figure}
\centerline{\includegraphics[width=8.3cm]{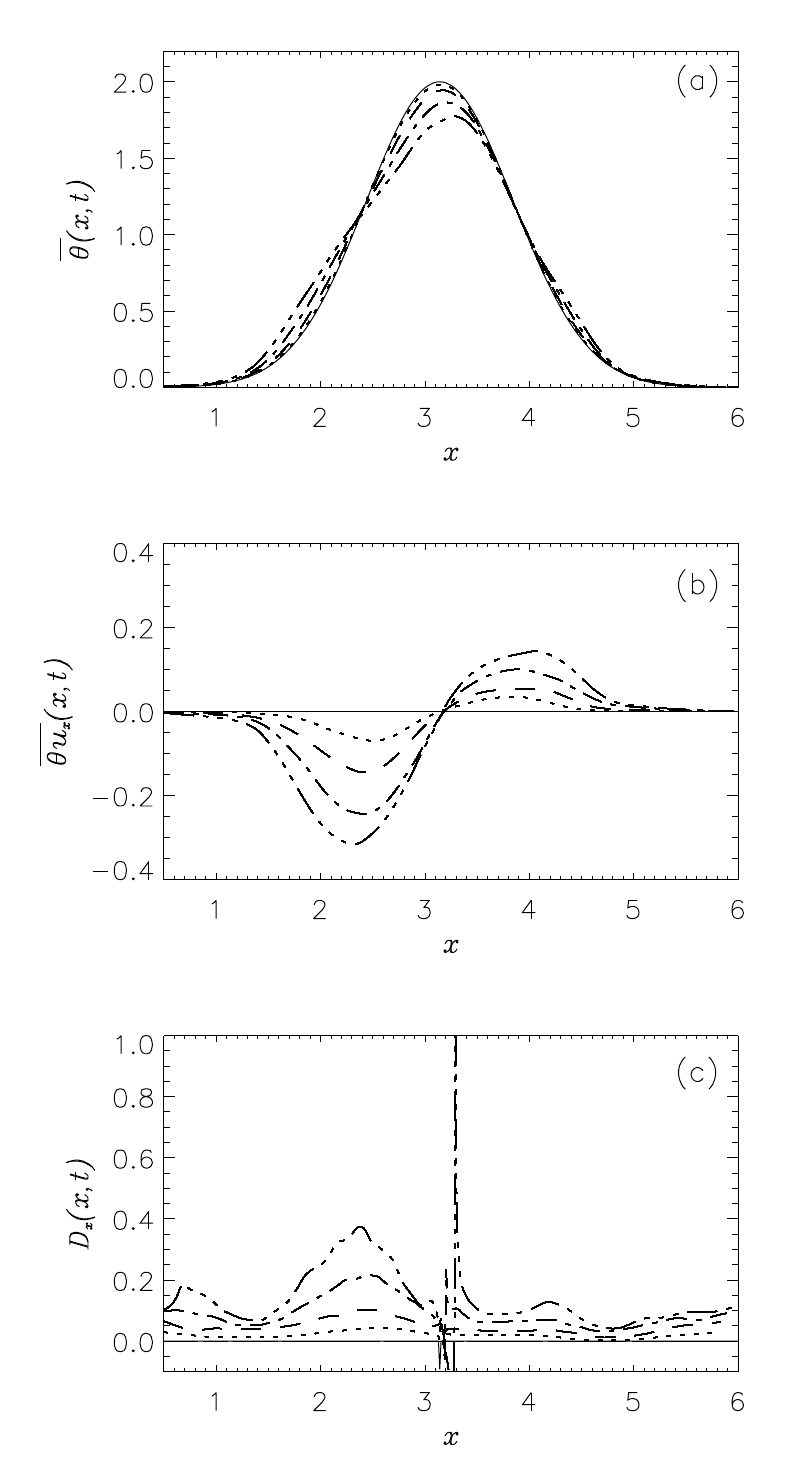}}
\caption{(a) Averaged horizontal concentration $\overline{\theta}$ in 
    run B$3_{x}$, at times $t =0$ (solid), $0.25$ (dotted), $0.5$
    (dashed), $0.75$ (dash-dotted), and $1$ (dash-triple-dotted).
    (b) Horizontal flux at the same times. (c) ${\cal D}_x(x,t)$ at
    the same times.}
\label{fig:fig11}
\end{figure}

From Fick's law, the actual turbulent diffusion coefficient is the 
average of ${\cal D}_{i}(x_i,t)$ over the coordinate $x_i$, and if the
system is in a turbulent steady state, over time. From
Eq.~(\ref{eq:DT}), we can define these averaged diffusion coefficients 
as follows. We can first average over the coordinate $x_i$ to obtain 
a time dependent turbulent diffusion,
\begin{equation}
{\cal D}_i(t) = \frac{1}{2 \pi} \int_0^{2\pi} {\cal D}_i(x_i,t) \mbox{d}x_i ,
\end{equation}
and we can further average over time, to obtain the mean turbulent 
diffusion
\begin{equation}
{\cal D}_i = \frac{1}{T} \int_{t_0}^{t_0+T} {\cal D}_i(t) \mbox{d}t .
\label{eq:taverage}
\end{equation}
Here, $t_0$ and $T$ are characteristic times of the flow. In practice, 
in our simulations the turbulent diffusion ${\cal D}_i(t)$ first grows
in time as the initial Gaussian profile is mixed by the turbulence,
then reaches an approximate steady state value for a few turnover 
times, and then decreases as the scalar becomes completely diluted 
(which happens after three or four turnover times).

\begin{figure}
\centerline{\includegraphics[width=8.5cm]{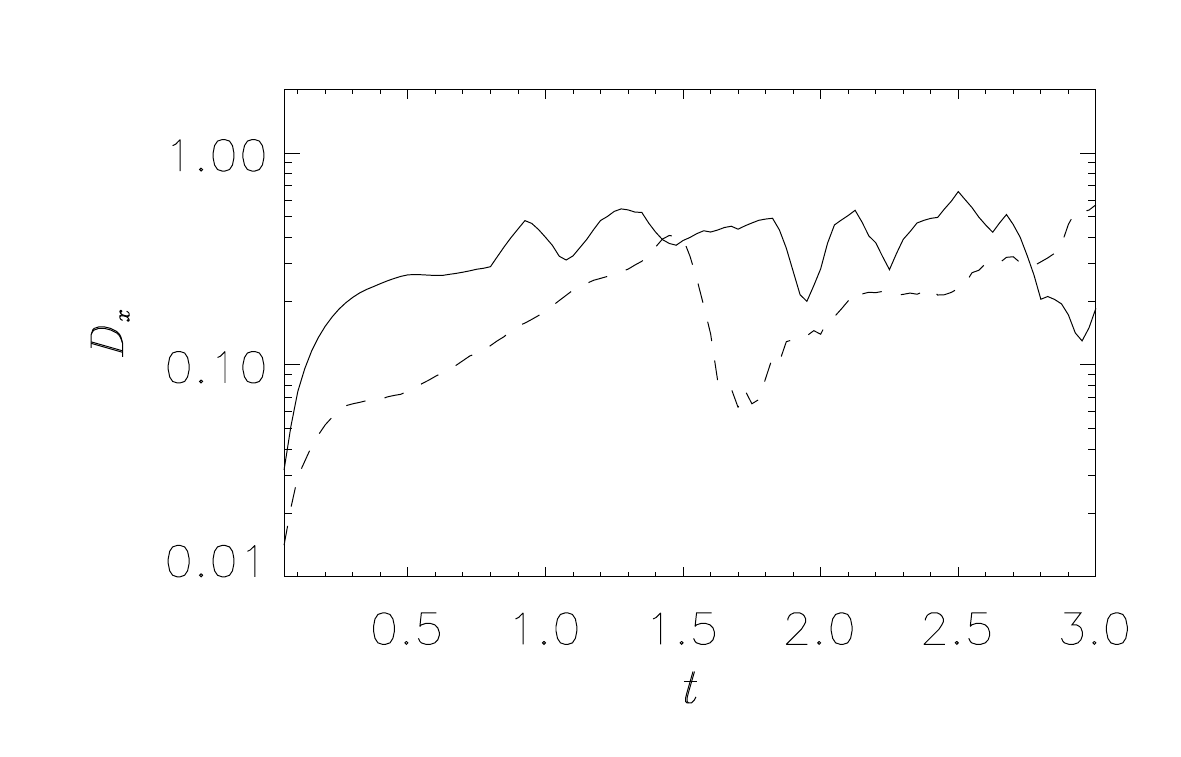}}
\caption{Horizontal turbulent diffusion as a function of time for 
runs A$2_{x}$ (solid) and B$2_{x}$ (dashed) ($\textrm{Ro} = 0.02$, 
respectively without and with helicity).}
\label{fig:fig12}
\end{figure}

\subsection{Isotropic helical turbulence}

In the absence of rotation, diffusion coeficients are expected to be 
isotropic, and therefore horizontal and vertical turbulent diffusion 
should be the same within error bars. Figure \ref{fig:fig8} shows 
the mean passive scalar profile $\overline{\theta}(x,t)$, the 
horizontal flux $\overline{\theta u _{x}}(x,t)$, and the pointwise
value of ${\cal D}_{x}(x,t)$, at five different times for run A$1_{x}$ 
(no rotation and no net helicity). 

As time evolves, the mean profile $\overline{\theta}(x,t)$ flattens 
and widens. The flux is antisymmetric: it is positive for $x > \pi$ 
and negative for $x < \pi$. This behavior for the flux is to be 
expected, as at $t = 0$ there is an excess of passive scalar 
concentration at $x = \pi$ that must be transported by turbulent 
diffusion towards $x = 0 $ and towards $x =  2\pi $. The pointwise
value of ${\cal D}_x(x,t)$ fluctuates around a mean value (which
increases with time), except close to $x = \pi$ where it rapidly 
takes very large positive and negative values as in that point 
$\partial _{x} \overline{\theta}$ approaches zero. The mean spatial
value of ${\cal D}_x(x,t)$ increases to its saturation value around
$t_{0} \approx 1.5$; after this time it fluctuates around its value 
(see more details below).

Figure \ref{fig:fig9} shows the same quantities at five different 
times for run B$1_{x}$ (i.e., in a simulation without rotation but 
with injection of net helicity). The behavior of the mean
concentration of the passive scalar, the horizontal scalar flux, and 
the pointwise value of ${\cal D}_x(x,t)$ is qualitatively the same as
in the non-helical run A$1_{x}$. However, the helical run displays a
larger diffusion of the mean concentration of the scalar (as 
evidenced by the smaller maximum value of $\overline{\theta}(x,t)$
around $x=\pi$ and by the stronger tails close to to $x=0$ and 
$2\pi$, when curves at the same time are compared in 
Figs.~\ref{fig:fig8} and \ref{fig:fig9}). Also, the spatial flux 
$\overline{\theta u_x}$ takes larger extreme values in the helical
simulation, and the spatial average of ${\cal D}_x(x,t)$ seems to
result in larger values for the turbulent diffusion in this run.

\begin{figure}
\centering
\includegraphics[width=8.5cm]{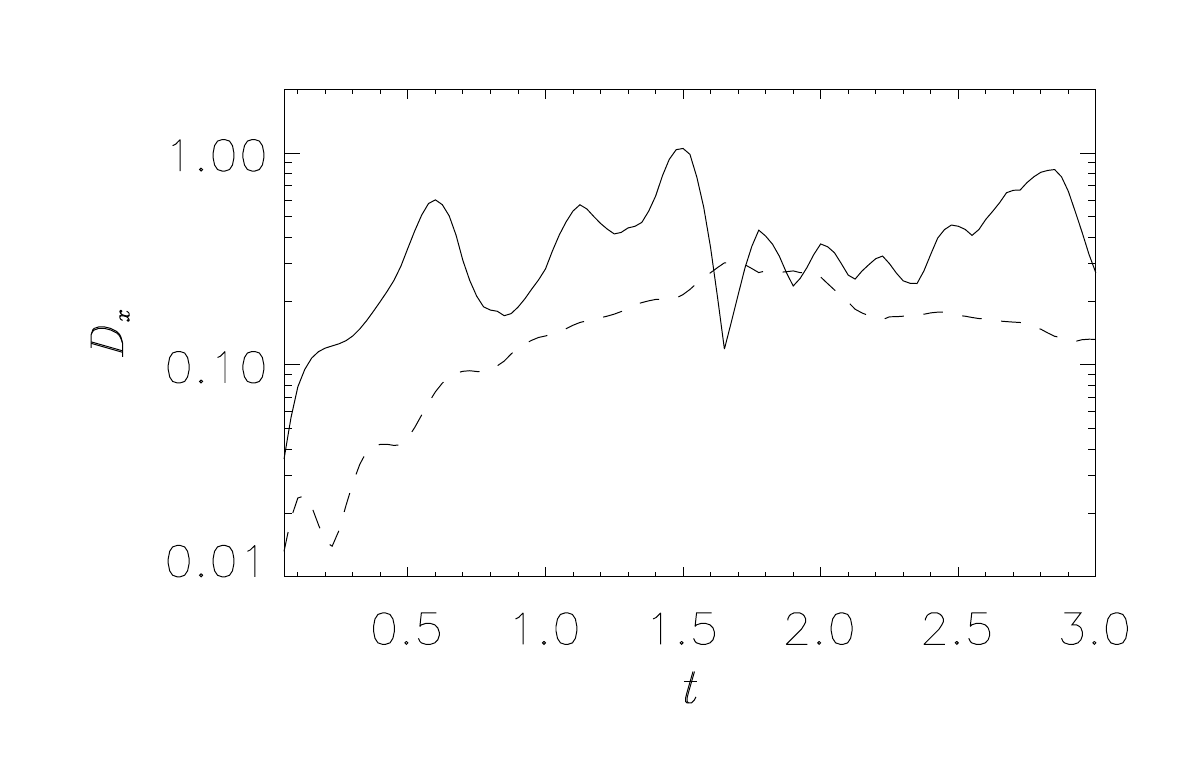}
\caption{Horizontal turbulent diffusion as a function of time for 
runs A$3_{x}$ (solid) and B$3_{x}$ (dashed) ($\textrm{Ro}= 0.01$, 
respectively without and with helicity).}
\label{fig:fig13}
\end{figure}

The increased turbulent diffusion in the presence of helicity is 
confirmed in Fig.~\ref{fig:fig10}, which shows the horizontal
turbulent diffusion as a function of time for runs A$1_{x}$ 
and B$1_{x}$. In both runs ${\cal D}_{x} (t)$ grows from an initially
small value to its saturation value around $t_{0} \approx 1.5$. As 
observed above, turbulent diffusion saturates at similar times for 
the helical and the non-helical case, but to a larger value in the 
presence of helicity.  Although in isotropic turbulence helicity does
not affect significantly the energy scaling 
\cite{chen03,chen03b,mininni09,proceeding2013}, an increase in 
the turbulent diffusion in the presence of helicity was predicted 
in \cite{Chkhetiani06}. Using renormalization group techniques, 
the authors estimated that turbulent diffusion in a helical flow 
can be up to a $50 \%$ larger than in a non-helical flow. In our 
simulations the averaged in time value of ${\cal D}_{x} (t)$ is 
$\approx 0.3$ for run A$1_x$, and $\approx 0.4$ for run B$1_x$, 
in reasonable agreement with the theoretical result.

It is worth mentioning that the same analysis was perfomed in
simulations A$1_{z}$ and B$1_{z}$ (i.e., the same runs but with an 
initial Gaussian profile in the $z$-direction). As expected from the
flow isotropy, the same behavior was obtained.

\subsection{Rotating helical turbulence}

\subsubsection{Horizontal diffusion}

\begin{figure}
\centering
\includegraphics[width=8.3cm]{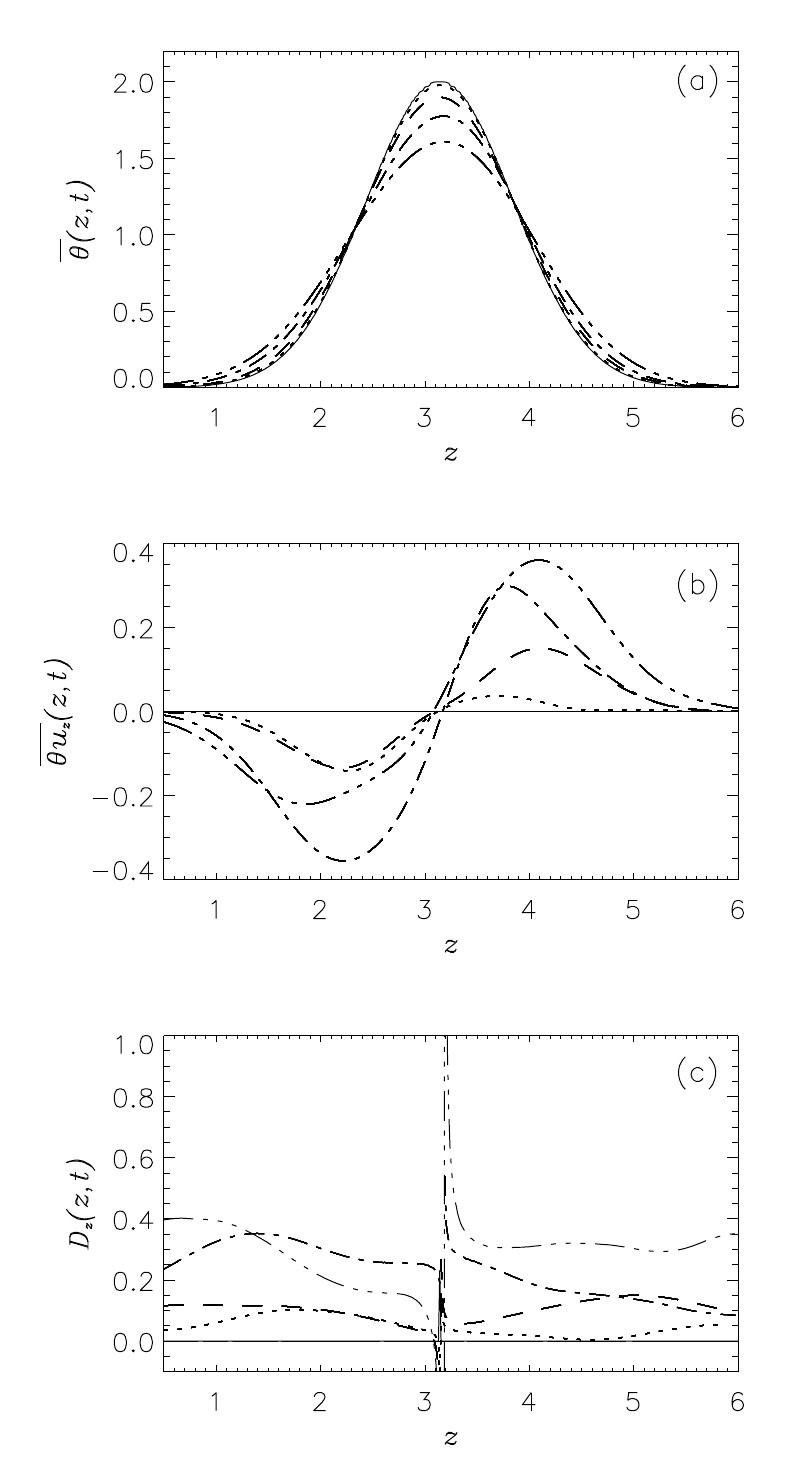}
\caption{Averaged vertical concentration $\overline{\theta}$ in run 
B$3_{z}$, at times $t =0$ (solid), $0.25$ (dotted), $0.5$ (dashed), 
$0.75$ (dash-dotted), and $1$ (dash-triple-dotted). (b) Horizontal 
flux at the same times. (c) ${\cal D}_x(x,t)$ at the same times.}
\label{fig:fig14}
\end{figure}

Figure \ref{fig:fig11} shows the mean profile of the passive scalar 
$\overline{\theta}(x,t)$, the horizontal flux 
$\overline{\theta u_{x}}(x,t)$, and the pointwise value of 
${\cal D}_{x}(x,t)$ at five different times for run B$3_{x}$ 
(helical and with $\textrm{Ro} = 0.01$). In this case, note that the 
average profile and the flux become asymmetric, i.e., there is an 
excess of concentration of $\overline{\theta}(x,t)$ for $x < \pi$, 
and the absolute value of the flux is larger for $x< \pi$ than for
$x > \pi$. This assymetry is caused by the Coriolis force and has 
been previously observed for rotating non-helical flows in 
\cite{Branden,imazio2013}. In our runs, the passive scalar at $t=0$ 
is concentrated in a narrow band around $x=\pi$. The average flux 
is thus towards positive values of $x$ for $x>\pi$ and towards 
negative values of $x$ for $x<\pi$ (i.e., in the direction of 
$-\nabla \theta$, see, for instance, Fig.~\ref{fig:fig16}). The 
Coriolis force in Eq.~(\ref{eq:NS}) is 
$-2 \Omega \hat{z} \times {\bf u}$, and creates an overturning in 
in the $x$-$y$ plane of the initially only dependent on $x$ 
Gaussian profile, as will be shown later in more detail in spatial
visualizations of the passive scalar. This overturning also results 
in the asymmetry in Fig.~\ref{fig:fig11} (for more details, see also 
\cite{imazio2013}).

By compiting the mean value of ${\cal D}_{x}(x,t)$ over the spatial 
coordinate we obtain the turbulent diffusion coefficient. Figure 
\ref{fig:fig12} shows first the horizontal turbulent diffusion as 
a function of time for runs A$2_{x}$ and B$2_{x}$ (both with 
$\textrm{Ro} = 0.02$, without and with helicity respectively), 
and then Fig.~\ref{fig:fig13} shows the same quantity for runs 
A$3_{x}$ and B$3_{x}$ ($\textrm{Ro} = 0.02$, without and with helicity 
respectively). For both rotation rates, we observe that horizontal 
diffusion is smaller in the presence of helicity. This result is the 
opposite to that observed for the isotropic runs in the previous 
section, for which helicity increased the turbulent diffusion. 

\begin{figure}
\centering
\includegraphics[width=8.3cm]{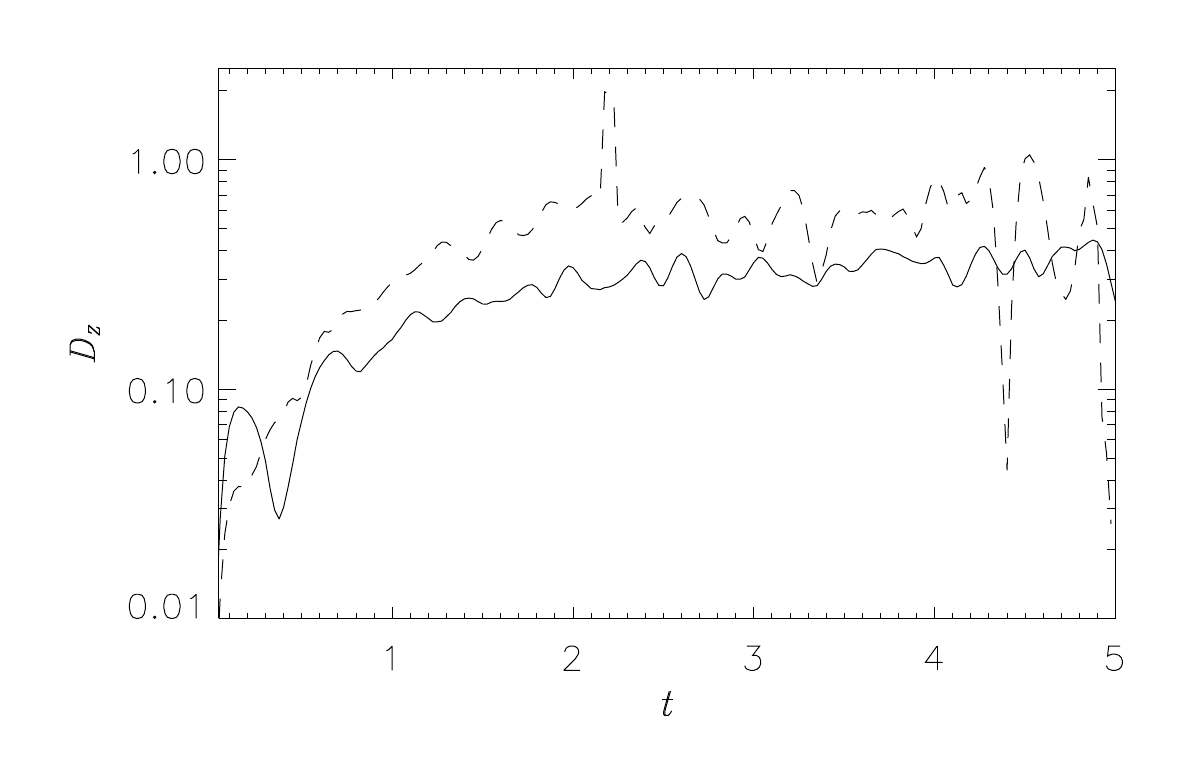}
\caption{Vertical turbulent diffusion as a function of time for 
runs A$3_{z}$ (solid) and $B3_{z}$ (dashed). The former run has 
no net helicity, while the latter has helical forcing.}
\label{fig:fig15}
\end{figure}

As already mentioned, while in isotropic turbulence helicity does 
not affect the energy spectrum scaling 
\cite{chen03,chen03b,mininni09,proceeding2013}, in rotating 
turbulence the presence of helicity results in shallower horizontal 
spectrum for the energy, in comparison with rotating non-helical 
turbulence \cite{mininni09,proceeding2013}. As a result, a smaller
turbulent diffusion can be expected, as small scale velocity field 
flutuations should be less energeting in the helical rotating case.
Indeed, in most two point closure models, the turbulent diffusivity 
is proportional to the mean kinetic energy in the turbulent 
fluctuations, $\overline{u^2}/2$, and if the kinetic energy spectrum 
is steeper, then the diffusivity should decrease. A simple mean 
field argument can illustrate this. We can split the velocity in a
mean flow $\overline{{\bf u}}$, and a fluctuating component 
${\bf u}'$, such that ${\bf u}=\overline{{\bf u}}+{\bf u}'$.
In our runs $\overline{{\bf u}}=0$, and ${\bf u}={\bf u}'$. Splitting 
the passive scalar in the same way we have 
$\theta=\overline{\theta}+\theta'$. Replacing in Eq.~(\ref{eq:theta}) 
and averaging we obtain
\begin{equation}
\frac{\partial \overline{\theta}}{\partial t}=-\nabla 
    \cdot(\overline{{\bf u}\theta'}),
\label{eq:17}
\end{equation}
and subtracting this equation from Eq.~(\ref{eq:theta}) we then obtain
\begin{equation}
\frac{\partial \theta}{\partial t}=-\nabla \cdot({\bf u}\overline{\theta}).
\end{equation}
We can integrate this last equation assuming the flow is correlated 
over the integral eddy turnover time $\tau$, to obtain 
\begin{equation}
\theta'\approx -\tau \nabla \cdot ({\bf u} \overline{\theta})=
    -\tau {\bf u} \cdot \nabla \overline{\theta},
\end{equation}
where incompressibility was used. Then, replacing in Eq.~(\ref{eq:17}),
\begin{equation}
\frac{\partial \overline{\theta}}{\partial t}\approx \frac{\partial}{\partial x_i}
\left(\tau \overline{u_iu_j}\right)\frac{\partial \overline{\theta}}{\partial x_i},
\end{equation}
where the coefficient $\tau \overline{u_iu_j}$ can be interpreted as a
turbulent diffusion. If the flow is isotropic, then 
${\cal D}\approx \tau \overline{u^2}$. A more refined mean field
derivation of this expression can be found in 
\cite{Elperin,Blakman,branden12}, while two point closure derivations
can be found in \cite{Kraichnan76,Herring}.

\begin{figure}
\centering
\includegraphics[width=4.5cm]{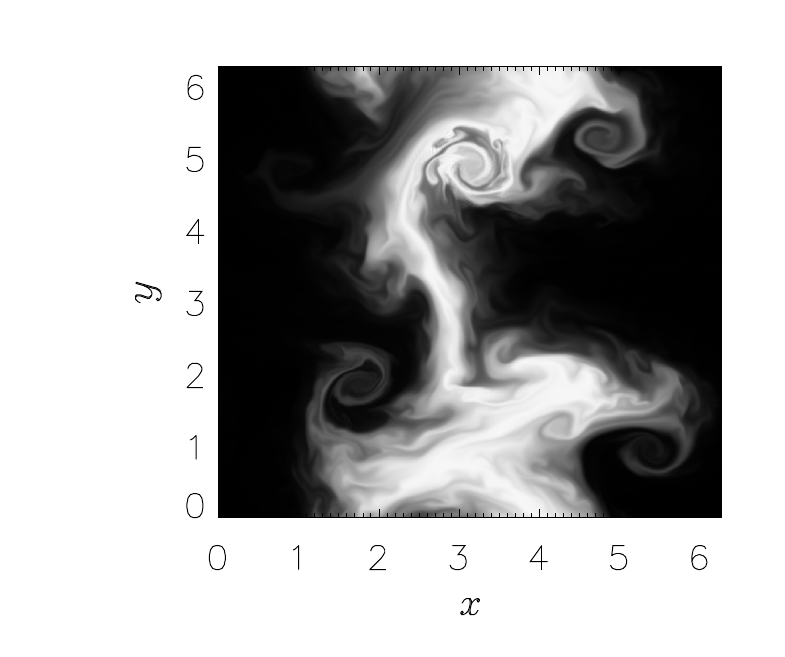}\includegraphics[width=4.5cm]{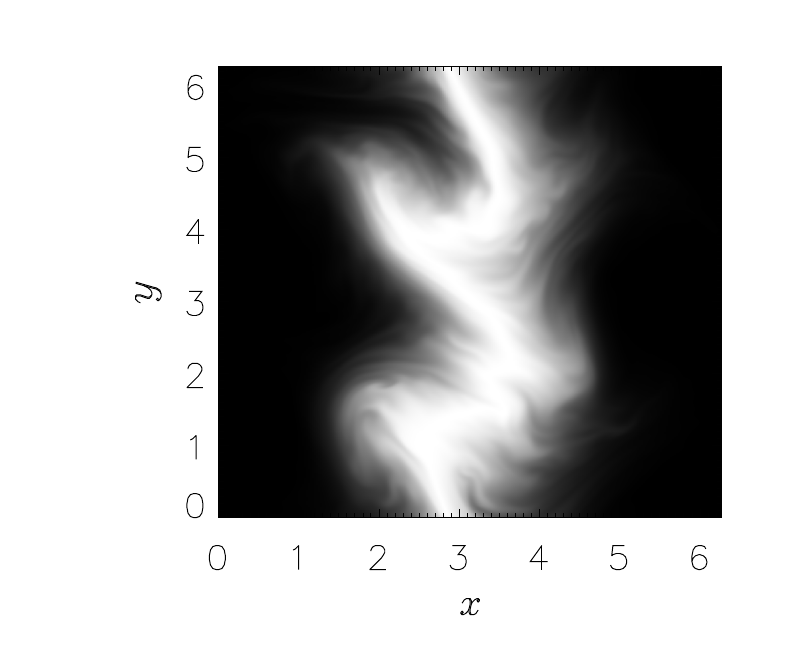} 
\caption{Passive scalar concentration in a horizontal slice of runs 
A$3_{x}$ (left) and $B3_{x}$ (right) at time $t = 1$. Note how the 
initial concentration (Gaussian, centered around $x=\pi$, and 
independent of the $y$-coordinate) gets distorted and diffussed.}
\label{fig:fig16}
\end{figure}

Although the argument above is only illustrative, it gives an
interesting hint to the possible cause of the reduced perpendicular
diffusion in helical rotating flows. As the perpendicular energy 
spectrum in this case is steeper than in the absence of helicity, 
then the smaller energy at small scales results in less mixing 
and diffusion.

\subsubsection{Vertical diffusion}

Figure \ref{fig:fig14} shows the mean vertical passive scalar 
concentration $\overline{\theta}(z)$, the mean vertical flux 
$\overline{\theta v_{z}}(z)$, and the pointwise value of 
${\cal D}_{z}(z)$ at different times in run B$3_{z}$. In this case,
the profiles are more similar to those obtained in the isotropic and
homogeneous case: $\overline{\theta}(z)$ and 
$\overline{\theta v_{z}}(z)$ are respectively symmetric and 
antisymmetric with respect to $z=\pi$. 

As in the case of horizontal diffusion, we can obtain the vertical 
turbulent diffusion coefficient as a function of time by computing 
the mean value of ${\cal D}_{z}(z,t)$ for all values of $z$. Figure 
\ref{fig:fig15} shows  ${\cal D}_{z}(t)$ for runs A$3_{z}$ and
B$3_{z}$ (both with $\textrm{Ro}=0.01$, respectively without and with
helicity). Note that horizontal turbulent diffusion is larger in the
presence of helicity, even more than in the isotropic case.

\subsection{Spatial distribution and structures}

Results shown above suggest that both horizontal and vertical 
diffusions are affected by rotation and by the presence of helicity. 
Figure \ref{fig:fig16} shows a horizontal slice of the passive scalar 
concentration in runs A$3_{x}$ and B$3_{x}$ at $t = 1$ (i.e., around
the time the turbulent diffusion coefficients ${\cal D}_x$ and 
${\cal D}_z$ reach a turbulent steady value). As also observed in 
\cite{imazio2013}, the initial Gaussian profile in the non-helical
rotating flow (run A$5_{x}$) diffuses in time, and also bends and 
rotates. As previously mentioned, the overturning of the profile is
caused by the Coriolis force (see also \cite{Branden}). In the helical
rotating flow (run B$3_{x}$), we also observe this overturning,
although the initial profile is less diffused (as indicated, e.g., by
the most extreme values in the $x$-coordinate for which a 
significant concentration of the passive scalar can be observed, which
are larger in run A$5_{x}$).

\begin{figure}
\centering
\includegraphics[width=4.5cm]{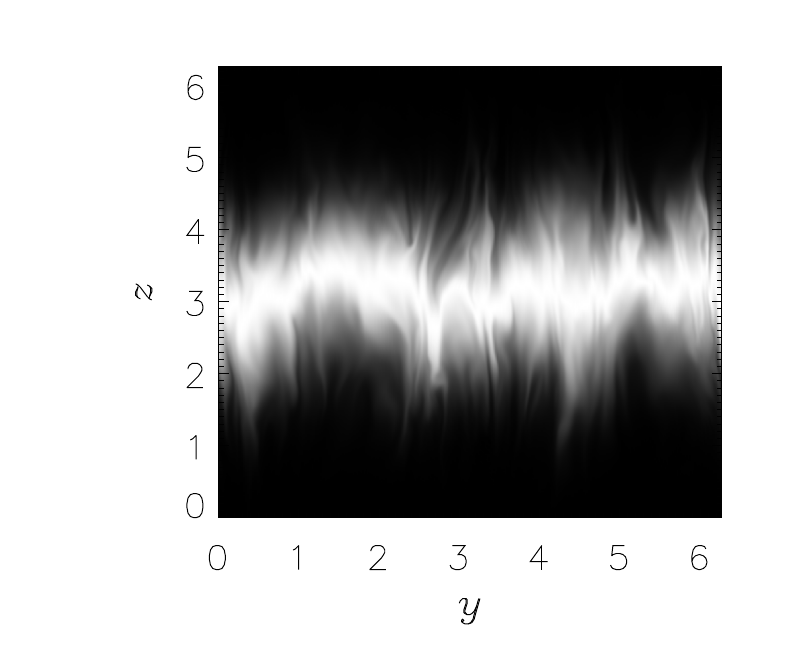}\includegraphics[width=4.5cm]{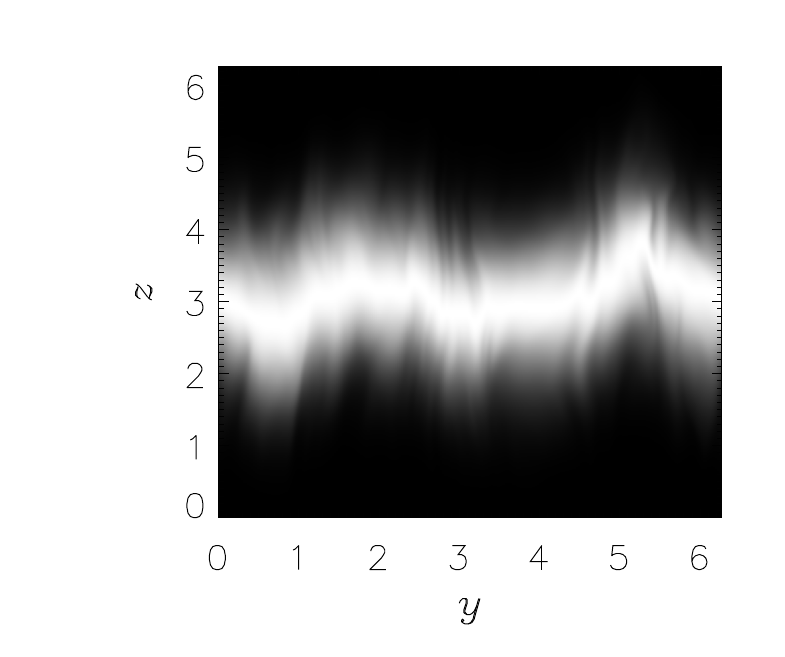}\\
\includegraphics[width=4.5cm]{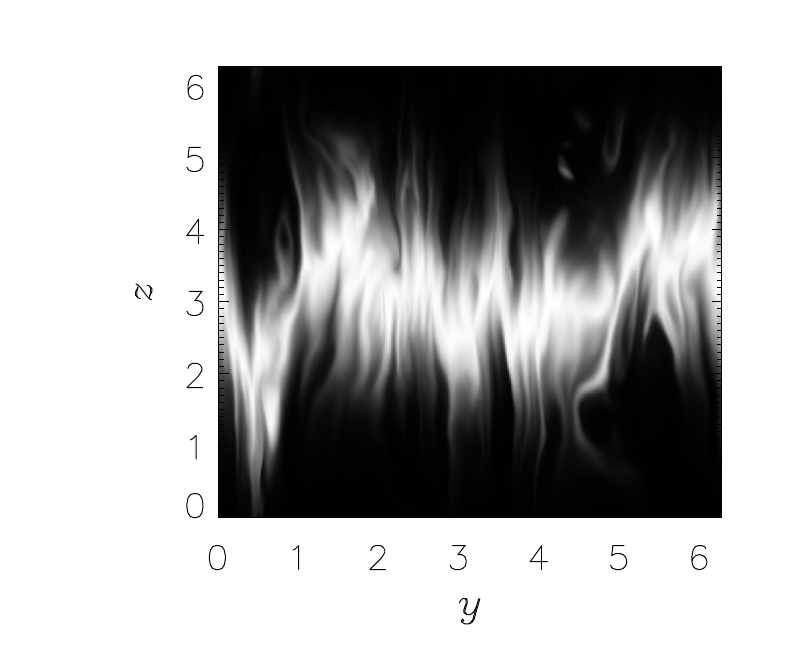}\includegraphics[width=4.5cm]{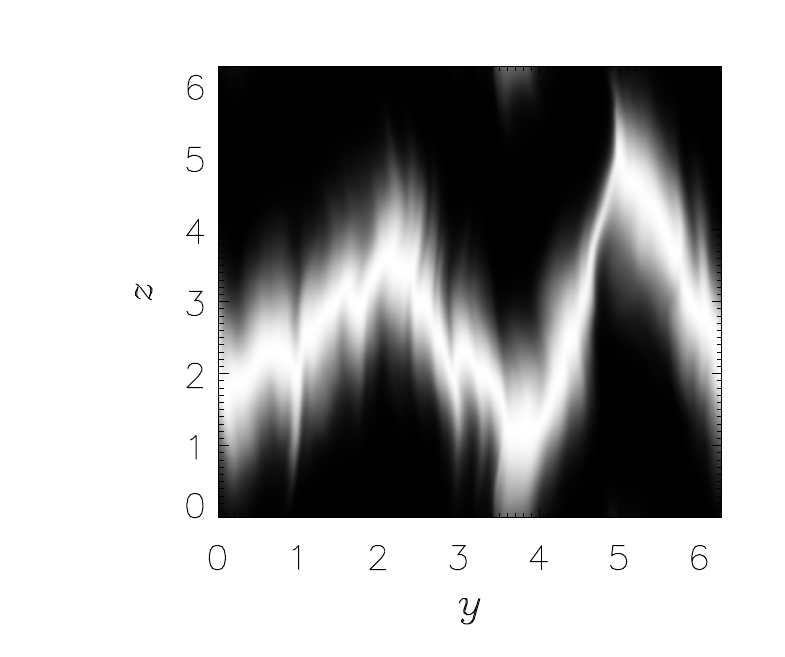} 
\caption{Passive scalar concentration in a vertical slice of runs 
A$3_{z}$ (left) and B$3_{z}$ (right) at time $t = 0.5$ (top row), and 
at time $t=1.5$ (bottom row).}
\label{fig:fig17}
\end{figure}

Diffusion in the parallel direction in rotating flows is of a
different nature than vertical diffusion (see Fig.~\ref{fig:fig17},
which shows vertical slices of the passive scalar concentration in
runs A$3_{z}$ and B$3_{z}$ at $t = 0.5$ and  $t = 1.5$.). In the
rotating non-helical case, the passive scalar initial profile is
diffused in vertical stripes, created by updrafts or downdrafts inside
columnar structures of the velocity field \cite{imazio2013}. These
columnar structures in the velocity and vorticity fields have been
reported in rapidly rotating flows, and are associated with the
bidimensionalization of the flow \cite{Cambon89,Waleffe,Davidson}. As
time increases, the stripes observed in the passive scalar in
Fig.~\ref{fig:fig17} are further streched, resulting in larger mixing
and diffusion. Note however that in the presence of helicity, the
stripes are increased even further, in good agrement with the
increased diffusion in helical flows reported above. This can be
understood as the presence of helicity in the flow requires the three
components of the velocity to be non-zero, resulting in a more
three-dimensional flow.

\section{Conclusions}

We analyzed data from direct numerical simulations of advection 
and diffusion of a passive scalar in rotating helical and non-helical 
turbulent flows. A total of $18$ simulations with spatial resolution 
of $512^3$ grid points was performed, using different Reynolds and 
Rossby numbers, and changing the forcing and initial conditions of the
passive scalar, to meassure energy and passive scalar spectra,
anisotropic velocity and passive scalar structure functions,
probability density functions, and diffusion coefficients in the 
directions parallel and perpendicular to the rotations axis.

In the first part of the paper we studied scaling laws of the energy 
and passive scalar variance, using spectra and structure functions in
the horizontal and vertical directions. We showed that helicity
affects the inertial range scaling of the passive scalar, with its
variance following a spectral law consistent with 
$\sim k_{\perp}^{-1.4}$. This scaling is shallower than the one found
for passive scalars in non-helical rotating turbulence
\cite{proceeding2013}, and consistent with a phenomenological 
argument that states that if the energy follows a power law 
$\sim k^{-n}$ in the inertial range, then the passive scalar
variance should follow a power law $\sim k^{-n_{\theta}}$ with
$n_{\theta}=(5-n)/2$. This argument, already proposed in 
\cite{proceeding2013} for rotating and non rotating non-helical 
flows, was found here to uphold also in the presence of helicity. 
The study of structure functions confirms these scaling laws, and 
indicates that the passive scalar is more anisotropic at small scales 
than velocity field. Also, the passive scalar was found to be more 
intermittent than the velocity field, a well known result, what which
becomes more pronounced in the presence of rotation and of helicity. 
The anomalous scaling exponents for the passive scalar can be
approximated using Kraichnan's model with the second order exponent
$\zeta_2$ obtained from our phenomenological model, and with a
dimensionality $d=2$. As in the case of rotating non-helical flows 
studied previously in \cite{imazio2011}, this value of $d$ was
interpreted as a result of the quasi-bidimentionalization of the
distribution of the passive scalar in the presence of rotation.

In the second part of the paper, the analisys of the effective 
diffusion coeficients calculated from Fick's law show that for 
isotropic flows (i.e., without rotation) helicity increases turbulent 
diffusion, in agreement with previous models and theoretical 
predictions \cite{Moffat83,Chkhetiani06}. In the presence of rotation, 
results indicate that the overall effect of rotation (irrespectively
of the content of helicity of the flow) is to decrease horizontal 
diffusion, while the effect on vertical diffusion is less pronounced. 
Helicity further decreases horizontal diffusion but increases vertical 
diffusion (compared with the non-helical rotating case). The decrease 
in horizontal diffusion was explained with a simple model for
turbulence diffusivity based on the available energy for the 
small-scale turbulent fluctuations.

\begin{acknowledgments}
The authors acknowledge support from grants No.~PIP 
11220090100825, UBACYT 20020130100738, PICT 2011-1529, 
and PICT 2011-1626. PDM acknowledges support from the Carrera 
del Investigador Cient\'{\i}fico of CONICET.
\end{acknowledgments}

\bibliography{ms}

\end{document}